\documentstyle[12pt]{article}
\begin{document}


\sloppy
\title
{\hfill{\normalsize\sf FIAN/TD/01-04}    \\
            \vspace{1cm}
{\Large The general form of the $*$--commutator on the Grassman algebra }
}
\author
 {
       I.V.Tyutin
          \thanks
             {E-mail: tyutin@td.lpi.ac.ru}
  \\
               {\small \phantom{uuu}}
  \\
           {\it {\small} I.E.Tamm Department of Theoretical Physics,}
  \\
               {\it {\small} P.N.Lebedev Physical Institute,}
  \\
         {\it {\small} 117924, Leninsky Prospect 53, Moscow, Russia.}
 }
\date{ }
\maketitle
\begin{abstract}
We study the general form of the $*$--commutator treated as a deformation of
the Poisson bracket on the Grassman algebra. We show that, up to a
similarity transformation, there are other deformations of the Poisson bracket
in addition to the Moyal commutator (one at even and one at odd $n$,
$n$ is the number of the generators of the Grassman algebra)
which are not reduced to the Moyal commutator by a similarity transformation.
\end{abstract}

\newpage

\section{Introduction}

As is well known, the major hopes for the construction of quantum
mechanics on nontrivial manifolds are connected with the so--called
geometric or deformation quantizations (\cite{BayFlaFroLicSte1} --
\cite{Kon}). The functions on phase space are put into correspondence with
the operators and the product of the operators and their commutator are
described by an associative $*$--product and $*$--commutator of
functions, which represent deformations of a usual ``pointwise'' product and
the Poisson bracket. On even manifolds, at least locally, the $*$--commutator
coincides with usual commutator in algebra with the associative $*$--product.
It is interesting to find out, what is the situation in the case, when the
phase space is a supermanifold. In present paper we investigate a general
form of the deformation of the nonsingular Poisson bracket on the Grassman
algebra. We show that on the Grassman algebra in addition to the
$*$--commutators which are equivalent to the Moyal commutator \cite{Moy},
there are also other deformations which are not reduced to the Moyal
commutator by a similarity transformation.

The paper is organized as follows. In Sect. 2 we formulate the problem.
In Sect. 3 the solution of the Jacobi identity considered as an equation to
the $*$--commutator is found in the lowest approximation in deformation
parameter. In Sect. 4 the higher deformations are considered and the main
result of the paper is formulated. In two Appendices solutions of the
equations considered in the main text are presented.

{\bf Notations and conventions.}

$\xi^\alpha$, $\alpha=1,\ldots,n$ are odd anticommuting generators
of the Grassman algebra:
$$
\varepsilon(\xi^\alpha)=1,\quad
\xi^\alpha\xi^\beta+\xi^\beta\xi^\alpha=0,\quad\partial_\alpha\equiv
{\partial\over\partial\xi^\alpha},
$$
$\varepsilon(A)$ denotes the Grassman parity of $A$;
$$
[\xi^\alpha]^0\equiv1,\quad[\xi^\alpha]^k\equiv\xi^{\alpha_1}
\cdots\xi^{\alpha_k}, 1\le k\le n,\quad[\xi^\alpha]^k=0, k>n,
$$
$$
[\partial_\alpha] ^0\equiv1, \quad [\partial_\alpha] ^k\equiv
\partial_{\alpha_1}\cdots\partial_{\alpha_k},1\le k\le n, \quad
[\partial_\alpha] ^k=0, k > n,
$$
$$
[\overleftarrow\partial_\alpha]^k\equiv\overleftarrow\partial_{\alpha_1}
\cdots\overleftarrow\partial_{\alpha_k},
$$
$$
T_{\ldots[\alpha]_k\ldots}\equiv T_{\ldots\alpha_1\ldots\alpha_k\ldots},
\quad T_{\ldots\alpha_i\alpha_{i+1}\ldots}=-
T_{\ldots\alpha_{i+1}\alpha_i\ldots},\quad i=1,\ldots,k-1.
$$

\section{Formulation of the problem}

We consider the Grassman algebra ${\cal G}_K$ over the field $K$ $=$
${\bf C}$ or ${\bf R}$, with the generators $\xi^\alpha$,
$\alpha=1,2,\ldots,n$. The generic element $f$ of the algebra (a function of
the generators) is
$$
f\equiv f(\xi)=\sum\limits_{k=0}^n{1\over k!}f_{[\alpha]_k}[\xi^\alpha] ^k,
\quad f_{[\alpha]_0}\equiv f_0=const,\,f_{[\alpha]_k}\in K.
$$

Let the nonsingular Poisson bracket
\begin{equation}\label{c1Poison}
\{f_1,f_2\}(\xi)\equiv[f_1,f_2]_{*_0}(\xi)\equiv c_0(\xi|f_1,f_2)=f_1(\xi)
\overleftarrow{\partial}_\alpha\omega^{\alpha\beta}(\xi)\partial_\beta
f_2(\xi)
\end {equation}
be given on the Grassman algebra ${\cal G}_K$ where
$\omega^{\alpha\beta}(\xi)$ is the symplectic metric, the even symmetric
tensor function satisfying the Jacobi identity:
$$
\omega^{\beta\alpha}=\omega^{\alpha\beta},\quad
\omega^{\alpha\delta}\partial_\delta\omega^{\beta\gamma}+
\omega^{\gamma\delta}\partial_\delta\omega^{\alpha\beta}+
\omega^{\beta\delta}\partial_\delta\omega^{\gamma\alpha}=0.
$$

We are interested in the general form of the $*$--commutator, i.e. the
bilinear mapping ${\cal G}_K\times{\cal G}_K$ $\rightarrow$ ${\cal G}_K$:
$$
f_3(\xi)\equiv[f_1,f_2]_*(\xi)=C(\xi|f_1,f_2),
$$
with the following properties:

i) The mapping is even
$$
 \varepsilon(C(\xi|f_1,f_2))=\varepsilon(f_1)+\varepsilon(f_2).
$$

ii) Antisymmetry
$$
C(\xi|f_2,f_1)=-(-1)^{\varepsilon(f_1)\varepsilon(f_2)}C(\xi|f_1,f_2).
$$

iii) The Jacobi identity
\begin{equation}\label{c1Jacob}
(-1)^{\varepsilon(f_1)\varepsilon(f_3)}[f_1,[f_2,f_3]_*]_*+
\hbox{cycle}(1,2,3)=0, \quad\forall f_1,f_2,f_3,
\end {equation}
or, equivalently,
\begin{equation}\label{c1Jacob'}
(-1)^{\varepsilon(f_1)\varepsilon(f_3)}C(\xi|f_1,f_{23})+
\hbox{cycle}(1,2,3)=0,\quad\forall f_1,f_2,f_3,
\end {equation}
where we denote $f_{ij}(\xi)\equiv C(\xi|f_i, f_j)$.

iv) It is supposed, that the $*$--commutator is defined in terms of the
series in the deformation parameter $\hbar^2$,
\begin {equation}\label{BC}
[f_1,f_2]_*(\xi)=\sum_{k=0}\hbar^{2k}[f_1,f_2]_{*_k}(\xi)=\sum_{k=0}
\hbar^{2k}c_k(\xi|f_1, f_2),
\end{equation}
where $[f_1,f_2]_{*_0}=c_0(\xi|f_1,f_2)$ coincides with the Poisson bracket
(\ref{c1Poison}) (the boundary condition or the correspondence principle).
Differently, we want to establish the general form of a deformation
of the Poisson bracket on the Grassman algebra. The deformation parameter is
denoted by $\hbar^2$ and, without loss of generality, it is possible to
consider $\hbar$ as real and positive number.

We treat condition (\ref{c1Jacob}) (or, equivalently, (\ref {c1Jacob'})) as
an equation on possible structure of the $*$--commutator. Note that $C_T$ (or,
equivalently, the $*_T$--commutator),
\begin {equation}\label{c1Sim}
C_T (\xi|f_1,f_2)=T^{-1}C(\xi|Tf_1,Tf_2),
\end{equation}
$T$ is a nonsingular even linear mapping ${\cal G}_K\rightarrow{\cal G}_K$,
satisfies the properties i) - iii), if $C(\xi|f_1,f_2)$ satisfies these
properties.  However, the boundary condition can change. In general, the
operator $T$ can be represented as
$$
T=T'(1 +\hbar^2t_1+\ldots),\quad T'=T\bigl|_{\hbar=0}.
$$
Let $T'=T_\eta$ be the operator responsible for the change of generators:
\begin{equation}\label{Chan}
T_\eta\xi^\alpha\!=\!\eta^\alpha(\xi),\,T_\eta^{-1}\xi^\alpha\!=\!
T_\zeta\xi^\alpha\!=\!\zeta^\alpha(\xi),\quad T_\eta f (\xi) =f(\eta(\xi)),
\,T_\eta^{-1}f(\xi)=T_\zeta f(\xi)=f(\zeta(\xi)),
\end{equation}
where $\zeta^\alpha(\xi)$ is inverse change, $\eta^\alpha(\zeta(\xi))=
\xi^\alpha$, $\varepsilon(\eta^\alpha)=\varepsilon(\zeta^\alpha)=1$. The
$*_{T_\eta}$--commutator obey the property iv) but now with modified Poisson
bracket:
$$
\omega^{\alpha\beta}(\xi)\quad\longrightarrow\quad
\omega^{\alpha\beta}_{T_\eta}(\xi)=\xi^\alpha
{\overleftarrow\partial\over\partial\zeta^{\alpha'}}\omega^{\alpha'\beta'}
(\zeta(\xi)){\partial\over\partial\zeta^{\beta'}}\xi^\beta.
$$
We say, that the $*_T$--commutator and $*$--commutator are related by a
similarity transformation, or call them equivalent. The $*$--commutator
obtained from the Poisson bracket by a similarity transformation is a trivial
deformation of the Poisson bracket.

We assume, that after appropriate similarity transformation the Poisson
bracket is of canonical form
\begin{equation}\label{c1Pois}
\{f_1,f_2\}(\xi)\equiv[f_1,f_2]_{*_0}(\xi)\equiv c_0(\xi|f_1,f_2)=f_1(\xi)
\overleftarrow{\partial}_\alpha\lambda_\alpha\partial_\alpha f_2 (\xi),\quad
\lambda^2_\alpha=1,
\end {equation}
where $\lambda_\alpha=1$ in the case of the Grassman algebra over complex numbers
(${\cal G}_{\bf C}$) and $\lambda_\alpha=\pm1$ in the case of the Grassman
algebra over real numbers (${\cal G}_{\bf R}$).

Eq. (\ref{c1Jacob}) (or equivalent eq. (\ref{c1Jacob'})) will be solved in
terms of expansion in deformation parameter $\hbar^2$. It is obviously
satisfied in the zeroth approximation. In the first order one gets
\begin{equation}\label{c1Jacob1}
(-1)^{\varepsilon(f_1)\varepsilon(f_3)}[f_1,[f_2,f_3]_{*_1}]_{*_0}+
(-1)^{\varepsilon(f_1)\varepsilon(f_3)}[f_1,[f_2,f_3]_{*_0}]_{*_1}+
\hbox{cycle}(1,2,3)=0,
\end {equation}
or, equivalently,
\begin{equation}\label{c1Jacob1'}
(-1)^{\varepsilon(f_1)\varepsilon(f_3)}[f_1,c_1(f_2,f_3)]_{*_0}(\xi)+
(-1)^{\varepsilon(f_1)\varepsilon(f_3)}c_1(\xi|f_1,[f_2,f_3]_{*_0})+
\hbox{cycle}(1,2,3)=0.
\end{equation}
Besides, the conditions i) and ii) should be satisfied, or
$$
c_1(\xi|f_2,f_1)=-(-1)^{\varepsilon(f_1)\varepsilon(f_2)}c_1(\xi|f_1,f_2),
\quad\varepsilon(c_1(\xi|f_1,f_2))=\varepsilon(f_1)+\varepsilon(f_2).
$$
The obvious solution of equation (\ref{c1Jacob1}) or (\ref{c1Jacob1'})
is a bilinear functional $c_{1triv} $, obtained from the Poisson
bracket by the similarity transformation $Tf=T(\xi|f)=(1+\hbar t_1+
O(\hbar^2))f(\xi)=f(\xi)+\hbar t_1(\xi|f)+O(\hbar^2)$,
$\varepsilon(Tf)=\varepsilon(t_1f)=\varepsilon(f)$:
\begin {equation}\label{c1Triv}
c_{1triv}(\xi|f_1,f_2)=[f_1,t_1f_2]_{*_0}(\xi)-t_1(\xi|[f_1,f_2]_{*_0})+
[t_1f_1,f_2]_{*_0}(\xi).
\end {equation}

Consider the algebra ${\cal B}=\sum\limits_\oplus{\cal B}_k$ of even
$k$--linear functionals $\Phi_k(\xi|f_1,\ldots,f_k)\in{\cal B}_k$
($k$--linear mappings $({\cal G}\times)^k\rightarrow{\cal G}$),
$\varepsilon(\Phi_k(\xi|f_1,\ldots,f_k))=\varepsilon(f_1)+\cdots+
\varepsilon(f_k)$ $(mod\,\,2)$, $k\ge1$, ${\cal B}_0=K$, with the property
$$
\Phi_k(\xi|f_q,\ldots,f_{i-1},f_i,f_{i+1},\ldots,f_n)=
-(-1)^{\varepsilon(f_i)\varepsilon(f_{i+1})}
\Phi_k(\xi|f_q,\ldots,f_{i-1},f_{i+1},f_i,\ldots,f_n).
$$
The product ($\circ$) in this algebra (a mapping
${\cal B}_k\times{\cal B}_l\rightarrow{\cal B}_{k+l}$) is defined as follows:
$$
\Phi_k\circ\Phi_l(\xi|f_1,\ldots,f_{k+l})={1\over(k+l)!}\sum_Psign(P)
\Phi_k(\xi|f_{i_1},\ldots,f_{i_k})\Phi_l(\xi|f_{i_{k+1}},\ldots,f_{i_{k+l}}),
$$
where the sum is taken over all permutations $P$ of numbers
$1,2,\ldots,k+l$; $sign(P)$ is the parity of the permutation $P$
calculated by the following rule:
the permutation of the neighboring indexes $i_p $ and $i_q $ gives the sign
factor $(-1)^{1+\varepsilon(f_{i_p})\varepsilon(f_{i_q})}$. In this algebra
there is a natural grading $g$: $g(\Phi_k)=k$, turning the algebra
${\cal B}$ into the graded algebra, $g({\cal B}_k)=k$, and the linear
operator $d_S$, the differential coboundary Chevalley operator
${\cal B}_k\rightarrow{\cal B}_{k+1}$, $g(d_H)=1$, acting according to the
rule:
$$
d_S\Phi_k(f_1,\ldots,f_{k+1})=\sum_{i=1}^{k+1}(-1)^{i+1+
\varepsilon(f_i)\sum\limits_{l=1}^{i-1}\varepsilon(f_l)}
f_i(\xi)*_0\Phi_k(\xi|f_1,\ldots,f_{i-1},f_{i+1},\ldots,f_{k+1})+
$$
$$
+\!\!\sum_{1\le i<j\le k+1}\!\!(-1)^{j+1+
\varepsilon(f_j)(\varepsilon(f_{i+1}+\cdots+\varepsilon(f_{j-1}))}
\Phi_k(\xi|f_1,\ldots,f_{i-1},f_i*_0f_j,f_{i+1},\ldots,f_{j-1},f_{j+1},
\ldots,f_{k+1}),
$$
$$
d_S{\cal B}_0=0.
$$
It is easy to prove that
$$
d_S^2=0,\quad d_S(\Phi_k\circ\Phi_l)=(d_S\Phi_k)\circ\Phi_l+
(-1)^{g(\Phi_k)}\Phi_k\circ d_S\Phi_l.
$$

Eqs. (\ref{c1Jacob1'}) and (\ref{c1Triv}) can be rewritten in terms of
the Chevalley operator $d_S$ as
$$
d_Sc_1(\xi|f_1,f_2,f_3)=0,
$$
$$
c_{1triv}(\xi|f_1,f_2)=d_St_1(\xi|f_1,f_2),
$$
i.e. it means that $c_1$ belongs to the second Chevalley cogomology group,
while the first order trivial deformations of the Poisson bracket are
coboundaries, i.e. they belong to the zeroth Chevalley cogomology. Thus, the
problem of description of all solutions of eq. (\ref{c1Jacob1'}) can
be now formulated as a problem of calculation of the second Chevalley
cogomology group $H^2(d_S,{\cal B})$.

\section {The first order deformation}

It is convenient to use the momentum representation of the $*$--commutator
(any bilinear functional can be represented in such a form)
$$
c_1(\xi|f_1,f_2)=f_1(\xi)\sum_{k,l=0}^n[\overleftarrow\partial_\alpha]^k
c_1^{[\alpha]_k|[\beta]_l}(\xi)[\partial_\beta]^lf_2(\xi).
$$

The properties i) and ii) give:
\begin{equation}\label{c2Asym}
\varepsilon(c_1^{[\alpha]_k|[\beta]_l})=k+l\,(mod\,\,2),\quad
c_1^{[\beta]_l|[\alpha]_k}=-(-1)^{kl}c_1^{[\alpha]_k|[\beta]_l}.
\end {equation}

The equation for the coefficient functions
$c_1^{[\alpha]_k|[\beta]_l}(\xi)$ follows from eq. (\ref{c1Jacob1'})
\begin{equation}\label{c2Jac}
\begin{array}{l}
\,\,\,\,f_1(\xi)\overleftarrow\partial_\gamma\lambda_\gamma
\partial_\gamma\biggl
(f_2(\xi)\sum\limits_{k,l=0}^n[\overleftarrow\partial_\alpha]^k
c_1^{[\alpha]_k|[\beta]_l}(\xi)[\partial_\beta]^lf_3(\xi)\biggr)+\\
+(-1)^{\varepsilon (f_2) \varepsilon (f_3)}
\biggl(f_1(\xi)\sum\limits_{k,l=0}^n[\overleftarrow\partial_\alpha]^k
c_1^{[\alpha]_k|[\beta]_l}(\xi)[\partial_\beta]^lf_3(\xi)\biggr)
\overleftarrow\partial_\gamma\lambda_\gamma\partial_\gamma f_2(\xi)- \\
-\biggl(f_1(\xi)\sum\limits_{k,l=0}^n[\overleftarrow\partial_\alpha]^k
c_1^{[\alpha]_k|[\beta]_l}(\xi)[\partial_\beta]^lf_2(\xi)\biggr)
\overleftarrow\partial_\gamma\lambda_\gamma\partial_\gamma f_3 (\xi)+ \\
+f_1(\xi)\sum\limits_{k,l=0}^n[\overleftarrow\partial_\alpha]^k
c_1^{[\alpha]_k|[\beta]_l}(\xi)[\partial_\beta]^l\biggl(f_2(\xi)
\overleftarrow\partial_\gamma\lambda_\gamma\partial_\gamma f_3(\xi)\biggr)+\\
+(-1)^{\varepsilon(f_2)\varepsilon(f_3)}
\biggl(f_1(\xi)\overleftarrow\partial_\gamma\lambda_\gamma\partial_\gamma
f_3(\xi)\biggr)\sum\limits_{k,l=0}^n[\overleftarrow\partial_\alpha]^k
c_1^{[\alpha]_k|[\beta]_l}(\xi)[\partial_\beta]^lf_2(\xi)-\\
-\biggl(f_1(\xi)\overleftarrow\partial_\gamma\lambda_\gamma\partial_\gamma
f_2(\xi)\biggr)\sum\limits_{k,l=0}^n[\overleftarrow\partial_\alpha]^k
c_1^{[\alpha]_k|[\beta]_l}(\xi)[\partial_\beta]^lf_3(\xi)=0.
\end{array}
\end{equation}

We also write the similarity transformation in the momentum representation
$$
t_1(\xi|f)=\sum_{k=0}^nt_1^{[\alpha]_k}(\xi)[\partial_\alpha]^kf(\xi),\quad
\varepsilon(t_1^{[\alpha]_k})=k\,(mod\,\,2),
$$
\begin{equation}\label{c2Triv}
\begin{array}{l}
\,\,\,\,\,c_{1triv}(\xi|f_1,f_2)=
f_1(\xi)\overleftarrow\partial_\gamma\lambda_\gamma\partial_\gamma\biggl
(\sum\limits_{k=0}^nt_1^{[\alpha]_k}(\xi)[\partial_\alpha]^kf_2(\xi)\biggr)-\\
-\sum\limits_{k=0}^nt_1^{[\alpha]_k}(\xi)[\partial_\alpha]^k
\biggl(f_1(\xi)\overleftarrow\partial_\gamma\lambda_\gamma\partial_\gamma
f_2(\xi)\biggr)+\biggl
(\sum\limits_{k=0}^nt_1^{[\alpha]_k}(\xi)[\partial_\alpha]^kf_1(\xi)\biggr)
\overleftarrow\partial_\gamma\lambda_\gamma\partial_\gamma f_2(\xi).
\end{array}
\end{equation}

We solve eq. (\ref {c2Jac}) studying the coefficients in front of
different degrees of the derivatives of the functions $f_i$. From now on it
is assumed that $n\ge2$.

{ \bf S t e p 0\quad} Consider the factor in eq. (\ref{c2Jac}) in front of
$f_1(\xi)$ (without derivatives):
$$
(-1)^{\varepsilon(f_2)\varepsilon (f_3)}
\biggl(\sum_{l=0}^nc_1^{0|[\beta]_l}(\xi)[\partial_\beta]^lf_3(\xi)\biggr)
\overleftarrow\partial_\gamma\lambda_\gamma\partial_\gamma f_2(\xi)-
$$
$$
-\biggl(\sum_{l=0}^nc_1^{0|[\beta]_l}(\xi)[\partial_\beta]^lf_2(\xi \biggr)
\overleftarrow\partial_\gamma\lambda_\gamma\partial_\gamma f_3(\xi)+
\sum_{l=0}^nc_1^{0|[\beta]_l}(\xi)[\partial_\beta]^l\biggl(f_2(\xi)
\overleftarrow\partial_\gamma\lambda_\gamma\partial_\gamma f_3(\xi)\biggr)
=0.
$$

From antisymmetry condition (\ref{c2Asym}) it follows that
$c_1^{0|0}(\xi)=0$.

${\bf 0_a}\quad$ The coefficient in front of $\partial_\alpha f_2(\xi)$ gives:
$$
\sum_{l=1}^n(-1)^lc_1^{0|[\beta]_l}(\xi)
\overleftarrow\partial_\alpha\lambda_\alpha[\partial_\beta]^lf_3(\xi)-
c_1^{0|\alpha}(\xi)\overleftarrow\partial_\gamma\lambda_\gamma
\partial_\gamma f_3 (\xi)=0.
$$
From these equations it follows (see Appendix 1) that:
\begin{equation}\label{0a}
\lambda_\alpha\partial_\alpha c_1^{0|\beta}(\xi)+
\lambda_\beta\partial_\beta c_1^{0|\alpha}(\xi)=0\quad\Longrightarrow\quad
c_1^{0|\alpha}(\xi)=\lambda_\alpha\partial_\alpha c_{10}{\xi},\quad
\varepsilon(c_{10})=0,
\end{equation}
$$
\partial_\alpha c_1^{0|[\beta]_l}(\xi)[\partial_\beta]^lf_3(\xi)=0
\quad\Longrightarrow\quad c_1^{0|[\beta]_l}(\xi)=const,\quad l\ge2.
$$

${\bf 0_b}\quad$ The coefficient in front of
$\partial_\alpha\partial_\beta f_2(\xi)$,
$f_3(\xi)=\exp{(\xi^\gamma p_\gamma)}$, $p_\alpha\xi^\beta+
\xi^\beta p_\alpha=p_\alpha p_\beta+p_\beta p_\alpha=0$, provides:
\begin{equation}\label{c20b}
\lambda_\alpha c_1^{0|[\gamma]_{l-1}\alpha}[p_\gamma]^{l-1}p_\beta-
\lambda_\beta c_1^{0|[\gamma]_{l-1}\beta}[p_\gamma]^{l-1}p_\alpha =0,\quad
l\ge2.
\end{equation}
The general solution of eq. (\ref {c20b}) is (see Appendix 1):
$$
c_1^{0|[\alpha]_l}=0,\,\,\,2\le l\le n-1,\quad
c_1^{0|[\alpha]_n}=c_{1n}\varepsilon^{[\alpha]_n},\quad
c_{1n}=const,\quad\varepsilon(c_{1n})=n\,(mod\,\,2).
$$

The contribution from the similarity transformation to the factor
$c_1^{0|[\alpha]_k}$ arises only from the term $t_1^0 $ in the expression
(\ref{c2Triv}) and is equal to
$$
c_{1triv}^{0|\alpha}=-\lambda_\alpha\partial_\alpha t_1^0(\xi),\quad
c_{1triv}^{0|[\alpha] _k}=0,\quad k\ge2.
$$
Performing the similarity transformation with $t_1^0 (\xi)=c_{10}(\xi)$, we
obtain the expression for $*$--commutator with
$c_1^{0|[\alpha]_k}=c_1^{[\alpha]_k|0}=0$, $k\le n-1$, and
$$
c_1^{0|[\alpha]_n}=c_{1n}\varepsilon^{[\alpha]_n},\quad
c_1^{[\alpha]_n|0}=-c_{1n}\varepsilon^{[\alpha]_n},\quad
c_{1n}=const,\quad\varepsilon(c_{1n})=n\,(mod\,\,2).
$$
Note, that in the case of the Grassman algebras over ${\bf C}$ or ${\bf R}$
the coefficient $c_{1n}$ can be different from zero only at even $n$.
Furthermore, if one requires, that the functions $f(\xi)=const$ $*$--commute
with any function, then $c_{1n}=0$.

{\bf S t e p 1\quad} The coefficient in eq. (\ref{c2Jac}) in front of
$\partial_\gamma f_1(\xi)$ gives:
\begin{equation}\label{c2Jac1}
\begin{array}{l}
\lambda_\gamma
f_2(\xi)\sum\limits_{k,l=1}^n(-1)^k[\overleftarrow\partial_\alpha]^k
\partial_\gamma c_1^{[\alpha]_k|[\beta]_l}(\xi)[\partial_\beta]^lf_3(\xi)-
\sum\limits_{l=1}^n\partial_\alpha
c_1^{\gamma|[\beta]_l}(\xi)[\partial_\beta]^lf_2(\xi)\lambda_\alpha
\partial_\alpha f_3(\xi)+ \\
+(-1)^{\varepsilon(f_2)}\lambda_\alpha\partial_\alpha f_2(\xi)
\sum\limits_{l=1}^n\partial_\alpha c_1^{\gamma|[\beta]_l}(\xi)
[\partial_\beta]^lf_3(\xi)-2c_ {1n}\varepsilon^ {[\beta] _n}
[\partial_\beta]^nf_2(\xi)\lambda_\gamma\partial_\gamma f_3(\xi)+ \\
+(-1)^{\varepsilon(f_2)}2\lambda_\gamma\partial_\gamma f_2 (\xi)c_{1n}
\varepsilon^{[\beta]_n}[\partial_\beta]^nf_3(\xi)-
\sum\limits_{l=0}^n\biggl
[c_1^{\gamma|[\beta]_l}[\partial_\beta]^l\biggl(\partial_\alpha f_2(\xi)
\lambda_\alpha\partial_\alpha f_3(\xi)\biggr)- \\
-c_1^{\gamma|[\beta]_l}[\partial_\beta]^l\partial_\alpha f_2(\xi)
\lambda_\alpha\partial_\alpha f_3(\xi)-
(-1)^{l(1+\varepsilon(f_2))}c_1^{\gamma|[\beta]_l}
\partial_\alpha f_2 (\xi) \lambda_\alpha
[\partial_\beta] ^l\partial_\alpha f_3 (\xi)\biggr]=0.
\end{array}
\end{equation}

${\bf 1_a}\quad$ The coefficient in front of $\partial_\alpha f_2(\xi)$,
$[\partial_\gamma] ^kf_3(\xi)$ provides:
$$
\lambda_\alpha\partial_\alpha c_1^{\beta|[\gamma]_k}(\xi)+
\lambda_\beta\partial_\beta c_1^{\alpha|[\gamma] _k}(\xi)+\delta_{1k}
\lambda_\gamma\partial_\gamma c_1^{\beta|\alpha}(\xi)+
\delta_{kn}2c_{1n}\lambda_\alpha\delta_{\alpha\beta}\varepsilon^{[\gamma]_n}
=0,\quad k=1,\ldots,n.
$$

${\bf 1_{a1}}\quad$ $k=1$
\begin{equation}\label{1a1}
\lambda_\alpha\partial_\alpha c_1^{\beta|\gamma}(\xi)+
\lambda_\beta\partial_\beta c_1^{\alpha|\gamma}(\xi)+
\lambda_\gamma\partial_\gamma c_1^{\beta|\alpha}(\xi)=0\quad
\Longrightarrow\quad
c_1^{\alpha|\beta}(\xi)=\lambda_\alpha\partial_\alpha c_1^\beta(\xi)+
\lambda_\beta\partial_\beta c_1^\alpha(\xi),
\end{equation}
with some function $c_1^\alpha(\xi)$ (see Appendix 1); here it was taken
into account that from antisymmetry properties (\ref{c2Asym}) of
the coefficients $c_1^{[\alpha]_k|[\beta]_l}$ it follows that
$c_1^{\alpha|\beta}=c_1^{\beta|\alpha}$.

${\bf 1_{a2}}\quad$ $2\le k\le n-1$
\begin {equation}\label{1a2}
\lambda_\alpha\partial_\alpha c_1^{\beta|[\gamma]_k}(\xi)+
\lambda_\beta\partial_\beta
c_1^{\alpha|[\gamma]_k}(\xi)=0,\quad\Longrightarrow\quad
c_1^{\alpha|[\beta]_k}(\xi)=\lambda_\alpha\partial_\alpha
c_1^{[\beta]_k}(\xi),\quad k=2,\ldots,n-1,
\end{equation}
with some functions $c_1^{[\beta]_k}(\xi)$ (see Appendix 1).

${\bf 1_{an}}\quad$ $k=n$
\begin {equation}\label{1an}
\lambda_\alpha\partial_\alpha c_1^{\beta|[\gamma]_n}(\xi)+
\lambda_\beta\partial_\beta c_1^{\alpha|[\gamma]_n}(\xi)=\lambda_\alpha
c_{1n}\delta_{\alpha\beta}\varepsilon^{[\gamma]_n}\quad\quad\Longrightarrow
\end{equation}
$$
c_1^{\alpha|[\beta]_n}(\xi)=
(\lambda_\alpha\partial_\alpha c'_{1n}(\xi)-{1\over2}\xi^\alpha
c_{1n})\varepsilon^{[\beta]_n},
$$
with some function $c'_{1n}(\xi)$. Perform the similarity transformation with
$t_1^0 (\xi)=0$, $t_1^{[\alpha]_k}(\xi)=-c_1^{[\alpha]_k}(\xi)$,
$k=1,\ldots,n-1$, $t_1^{[\alpha]_n}=-c'_{1n}(\xi)\varepsilon^{[\alpha]_n}$.
It does not change the coefficients $c^{0|[\alpha]_k}$ and reduces
$c_1^{\alpha|[\beta]_k}$ to the form
$$
c_1^{\alpha|[\beta]_k}(\xi)=0,\quad k\le n-1,\quad
c_1^{\alpha|[\beta]_n}(\xi)=-{1\over2}\xi^\alpha
c_{1n}\varepsilon^{[\beta]_n}.
$$

${\bf 1_b}\quad$ The coefficients in front of
$[\partial_\alpha]^kf_2(\xi)[\partial_\beta]^lf_3(\xi)$, $k,l\ge2$, give:
\begin{equation}\label{c21kl}
\begin{array}{l}
\,\,\,\,\,\lambda_\gamma f_2(\xi)[\overleftarrow\partial_\alpha]^k
\partial_\gamma c_1^{[\alpha]_k|[\beta]_l}(\xi)[\partial_\beta]^lf_3(\xi)-\\
-\delta_{k+l,n+2}(-1)^{(l-1)\varepsilon(f_2)}C_n^{k-1}q\xi^\gamma
c_{1n}\varepsilon^{[\alpha]_{k-1}[\beta]_{l-1}}\lambda_\sigma
\partial_\sigma[\partial_\alpha]^{k-1}f_2(\xi)
\partial_\sigma[\partial_\beta]^{l-1}f_3(\xi)=0.
\end{array}
\end{equation}
It is easy to see that second term in eq. (\ref{c21kl}) is identically
equal to zero, so from eq. (\ref {c21kl}) it follows
$$
c_1^{[\alpha]_k|[\beta]_l}(\xi)=c_1^{[\alpha]_k|[\beta]_l}=const,\quad
\forall\,\,k,l\ge2.
$$

For $n=2$ we have $c_1^{[\alpha]_2|[\beta]_2}=
c_1^{(2)}\varepsilon^{[\alpha]_2}\varepsilon^{[\beta]_2}$. Antisymmetry
condition (\ref{c2Asym}) gives $c_1^{(2)}=-c_1^{(2)}=0$.  Thus one obtains
for $n=2$
\begin{equation}\label{c2Gen12}
\begin{array}{l}\,\,\,\,\,[f_1,f_2]_{*_1}(\xi)=d_St_1(\xi|f_1,f_2)+\\
+c_{12}\biggl(f_1(\xi)(1-{1\over2}\overleftarrow\partial_\alpha\xi^\alpha)
\varepsilon^{[\beta]_2}[\partial_\beta]^2f_2(\xi)f_1(\xi)
[\overleftarrow\partial_\alpha]^2\varepsilon^{[\alpha]_2}
(1-{1\over2}\xi^\beta\partial_\beta)f_2(\xi)\biggr).
\end{array}
\end{equation}

In what follows $n\ge3$.

{\bf S t e p 2\quad} The coefficients in eq. (\ref{c2Jac}) in front of
$\partial_\alpha\partial_\beta f_1(\xi)$ and derivatives of $f_2(\xi)$ and
$f_3(\xi)$ of orders $\ge2$ give:
\begin{equation}\label{c2Jac2}
\begin{array}{l}
\,\,\,\,\,k\biggl(\lambda_\alpha f_2(\xi)
\overleftarrow\partial_\alpha[\overleftarrow\partial_\gamma]^{k-1}
c_1^{\beta[\gamma]_{k-1}|[\delta]_l}-\lambda_\beta f_2(\xi)
\overleftarrow\partial_\beta[\overleftarrow\partial_\gamma]^{k-1}
c_1^{\alpha[\gamma]_{k-1}|[\delta]_l}\biggr)[\partial_\delta] ^lf_3(\xi)+ \\
+lf_2(\xi)[\overleftarrow\partial_\gamma]^k
\biggl(c_1^{[\gamma]_k|\beta[\delta]_{l-1}}
\lambda_\alpha\partial_\alpha[\partial_\gamma]^{l-1}f_3(\xi)-
c_1^{[\gamma]_k|\alpha[\delta]_{l-1}}
\lambda_\beta\partial_\beta[\partial_\gamma]^{l-1}f_3(\xi)\biggr)- \\
-2(-1)^{k+kl}C^{k-1}_{k+l-2}\lambda_\sigma
f_2(\xi)\overleftarrow\partial_\sigma[\overleftarrow\partial_\gamma]^{k-1}
c_1^{\alpha\beta|[\gamma]_{k-1}[\delta]_{l-1}}
\partial_\sigma[\partial_\delta]^{l-1}f_3(\xi)=0,\quad k,l\ge2.
\end{array}
\end{equation}

${\bf 2_a}\quad$ $f_2(\xi)=\xi^\gamma\xi^\sigma $,
$f_3(\xi)=\exp{(\xi^\delta p_\delta)}$:
\begin{equation}\label{c22a}
\begin{array}{l}
\,\,\,\,\,\biggl(\delta_{\gamma\beta}\lambda_\beta
c_1^{\alpha\sigma|[\delta]_l}-\delta_{\gamma\alpha}\lambda_\alpha
c_1^{\beta\sigma|[\delta]_l}-\delta_{\sigma\beta}\lambda_\beta
c_1^{\alpha\gamma|[\delta]_l}+\delta_{\sigma\alpha}\lambda_\alpha
c_1^{\beta\gamma|[\delta]_l}\biggr)[p_\delta]^l= \\
=l\biggl(\delta_{\alpha\delta}c_1^{\gamma\sigma|\beta[\delta]_{l-1}}-
\delta_{\beta\delta}c_1^{\gamma\sigma|\alpha[\delta]_{l-1}}-
\delta_{\gamma\delta}c_1^{\alpha\beta|\sigma[\delta]_{l-1}}+
\delta_{\sigma\delta}c_1^{\alpha\beta|\gamma[\delta]_{l-1}}\biggr)
\lambda_\delta p_\delta[p_\delta]^{l-1}.
\end{array}
\end{equation}
The general solution of eq. (\ref{c22a}) is (see Appendix 1):
\begin{equation}\label{c22a'}
c_1^{\alpha\beta|[\delta]_l}[p_\delta]^l=
(\delta_{\alpha\delta}a_1^{\beta[\delta]_{l-1}}-\delta_{\beta\delta}
a_1^{\alpha[\delta]_{l-1}})\lambda_\delta p_\delta [p_\delta]^{l-1},\quad
l\ge2,
\end {equation}
where tensor $a_1^{\alpha_1\cdots\alpha_l}=a_1^{[\alpha]_l}$ is totally
antisymmetric. Note that according to formula (\ref{c22a'}),
$c_1^{\alpha\beta|[\delta]_n}=0$.

Perform the similarity transformation with
$t^{[\alpha]_k}_1(\xi)=(2/k)a_1^{[\alpha]_k}$, $k\ge2 $.
According to eq. (\ref{c2Triv}) one gets
$$
c_{1triv}(\xi|f_1,f_2)=-f_1(\xi)\overleftarrow\partial_\alpha
\overleftarrow\partial_\beta\sum_{l=2}(\delta_{\alpha\delta}
a_1^{\beta[\delta]_{l-1}}-\delta_{\beta\delta}
a_1^{\alpha[\delta]_{l-1}})\lambda_\delta
\partial_\delta[\partial_\delta]^{l-1}f_2(\xi)+\ldots,
$$
where dots mean the terms containing the third and higher derivatives of
the function $f_1(\xi)$. This transformation cancels the contribution to the
$*$--commutator containing the terms $c_1^{\alpha\beta|[\delta]_l}$,
$2\le l<n$, i.e., one may put
$$
c_1^{\alpha\beta|[\delta]_l}=0,\quad 2\le l\le n.
$$

{\bf S t e p 3\quad} Now eq. (\ref{c2Jac2}) is simplified because the
third term in the left hand side is equal to zero (including the cases
$k+l\ge n+2$):
\begin{equation}\label{c2Jac3}
\begin{array}{l}
\,\,\,\, \,k\biggl(\lambda_\alpha f_2(\xi)
\overleftarrow\partial_\alpha[\overleftarrow\partial_\gamma]^{k-1}
c_1^{\beta[\gamma]_{k-1}|[\delta]_l}-\lambda_\beta f_2(\xi)
\overleftarrow\partial_\beta[\overleftarrow\partial_\gamma]^{k-1}
c_1^{\alpha[\gamma]_{k-1}|[\delta]_l}\biggr)[\partial_\delta] ^lf_3(\xi)+ \\
+lf_2(\xi)[\overleftarrow\partial_\gamma]^k
\biggl(c_1^{[\gamma]_k|\beta[\delta]_{l-1}}
\lambda_\alpha\partial_\alpha[\partial_\delta]^{l-1}f_3(\xi)-
c_1^{[\gamma]_k|\alpha[\delta]_{l-1}}
\lambda_\beta\partial_\beta[\partial_\delta]^{l-1}f_3(\xi)\biggr)\!\!=0,\,
k,l\ge2.
\end{array}
\end{equation}

${\bf 3_a\quad}$ The coefficients in front of $[\partial_\gamma]^nf_2(\xi)$,
$f_3(\xi)=\exp{(\xi^\delta p_\delta)}$.

Introducing the notation $c_1^{[\alpha]_k|[\beta]_n}=c_1^{[\alpha]_k}
\varepsilon^{[\beta]_n}$, $c_1^{[\alpha]_n|[\beta]_l}=-(-1)^{ln}
c_1^{[\beta]_l}\varepsilon^{[\alpha] _n}$ and using the relation
$$
\overleftarrow\partial_\alpha[\overleftarrow\partial_\gamma]^{n-1}
\varepsilon^{\beta[\gamma]_{n-1}}={1\over n}\delta_{\alpha\beta}
[\overleftarrow\partial_\gamma]^n \varepsilon^{[\gamma] _n},
$$
we obtain ( the first term in eq. (\ref{c2Jac3}) is identically
equal to zero)
\begin{equation}\label{c23a1}
c_1^{\alpha[\delta]_{l-1}}\lambda_\beta p_\beta[p_\delta]^{l-1}=
c_1^{\beta[\delta]_{l-1}}\lambda_\alpha p_\alpha[p_\delta]^{l-1}.
\end{equation}

From eq. (\ref{c23a1}) it follows that (see Appendix 1) $c_1^{[\alpha]_k}=0$,
$k<n$, i.e.
$$
c_1^{[\alpha]_k|[\beta]_n}=c_1^{[\alpha]_n|[\beta]_l}=0,\quad k,l<n.
$$
For $k=l=n $ one has:
\begin{equation}\label{c2nn}
c_1^{[\alpha]_n|[\beta]_n}=\varepsilon^{[\alpha]_n}c_1^{(n)}
\varepsilon^{[\beta]_n},\quad \varepsilon(c_1^{(n)})=0,\quad
c_1^{(n)}=-(-1)^nc_1^{(n)},
\end {equation}
the last equality in (\ref {c2nn}) follows from antisymmetry condition
(\ref{c2Asym}) and that means that the coefficients $c_1^{(n)}$ can be
nonzero only for odd $n$.

It is useful to point out that the operator
$[\overleftarrow\partial_\alpha]^n\varepsilon^{[\alpha]_n}c_1^{(n)}
\varepsilon^{[\beta]_n}[\partial_\beta]^n$ can be also represented as
\begin{equation}\label{c2n3}
[\overleftarrow\partial_\alpha]^n\varepsilon^{[\alpha]_n}c_1^{(n)}
\varepsilon^{[\beta]_n}[\partial_\beta]^n=
c_1^{\prime(n)}\biggl
(\overleftarrow\partial_\alpha\lambda_\alpha\partial_\alpha\biggr)^n,\quad
c_1^{\prime(n)}=(-1)^pn!\lambda_1\cdots\lambda_nc_1^{(n)}.
\end {equation}

${\bf 3_b\quad}$ $k=l=3$
($f_2(\xi)=\xi^{\alpha_3}\xi^{\alpha_2}\xi^{\alpha_1}$,
$f_3(\xi)=\xi^{\beta_1}\xi^{\beta_2}\xi^{\beta_3}$).
\begin{equation}\label{c23b33}
\begin{array}{l}
\,\,\,\,\,\delta_{\alpha\alpha_1}\lambda_\alpha
c_1^{\beta\alpha_2\alpha_3|\beta_1\beta_2\beta_3}+\delta_{\alpha\alpha_2}
\lambda_\alpha c_1^{\beta\alpha_3\alpha_1|\beta_1\beta_2\beta_3}+
\delta_{\alpha\alpha_3}\lambda_\alpha
c_1^{\beta\alpha_1\alpha_2|\beta_1\beta_2\beta_3}- \\
-\delta_{\beta\alpha_1}\lambda_\beta
c_1^{\alpha\alpha_2\alpha_3|\beta_1\beta_2\beta_3}-\delta_{\beta\alpha_2}
\lambda_\beta c_1^{\alpha\alpha_3\alpha_1|\beta_1\beta_2\beta_3}-
\delta_{\beta\alpha_3}\lambda_\beta
c_1^{\alpha\alpha_1\alpha_2|\beta_1\beta_2\beta_3}+ \\
+\delta_{\alpha\beta_1} \lambda_\alpha
c_1^{\alpha_1\alpha_2\alpha_3|\beta\beta_2\beta_3}+\delta_{\alpha\beta_2}
\lambda_\alpha c_1^{\alpha_1\alpha_2\alpha_3|\beta\beta_3\beta_1}+
\delta_{\alpha\beta_3}\lambda_\alpha
c_1^{\alpha_1\alpha_2\alpha_3|\beta\beta_1\beta_2}- \\
-\delta_{\beta\beta_1}\lambda_\beta
c_1^{\alpha_1\alpha_2\alpha_3|\alpha\beta_2\beta_3}-\delta_{\beta\beta_2}
\lambda_\beta c_1^{\alpha_1\alpha_2\alpha_3|\alpha\beta_3\beta_1}-
\delta_{\beta\beta_3}\lambda_\beta
c_1^{\alpha_1\alpha_2\alpha_3|\alpha\beta_1\beta_2}=0.
\end{array}
\end{equation}
The general solution of eq. (\ref{c23b33}) is (see Appendix 1)
\begin{equation}\label{c23b334}
\begin{array}{l}
\,\,\,\,\,c_1^{\alpha_1\alpha_2\alpha_3|\beta_1\beta_2\beta_3}=
{\lambda_{\alpha_1}\lambda_{\alpha_2}\lambda_{\alpha_3}\Sigma\over
n(n-1)(n-2)}\biggl
(\delta_{\alpha_1\beta_1}\delta_{\alpha_2\beta_2}\delta_{\alpha_3\beta_3}+
\delta_{\alpha_1\beta_3}\delta_{\alpha_2\beta_1}\delta_{\alpha_3\beta_2}+
\delta_{\alpha_1\beta_2}\delta_{\alpha_2\beta_3}\delta_{\alpha_3\beta_1}- \\
-\delta_{\alpha_1\beta_2}\delta_{\alpha_2\beta_1}\delta_{\alpha_3\beta_3}-
\delta_{\alpha_1\beta_3}\delta_{\alpha_2\beta_2}\delta_{\alpha_3\beta_1}-
\delta_{\alpha_1\beta_1}\delta_{\alpha_2\beta_3}\delta_{\alpha_3\beta_2}
\biggr)
\end{array}
\end{equation}
or
\begin{equation}\label{c23b334'}
[\overleftarrow\partial_\alpha]^3c_1^{[\alpha]_3|[\beta]_3}[\partial_\beta]^3
={6\Sigma\over n(n-1)(n-2)}\biggl
(\overleftarrow\partial_\alpha\lambda_\alpha\partial_\alpha\biggr)^3.
\end{equation}
(Compare with expression (\ref{c2n3})).

${\bf 3_c\quad}$ We finally consider the remaining part of eq. (\ref{c2Jac3}).
Choosing $f_2(\xi)=\exp{(q_\gamma\xi^\gamma)}$, $f_3(\xi)=
\exp{(\xi^\delta p_\delta)}$, $q_\alpha q^\beta+q^\beta q_\alpha=p_\alpha
q_\beta+p_\beta q_\alpha=q_\alpha\xi ^\beta+\xi^\beta q_\alpha=0$, we have:
\begin{equation}\label{c23c}
\begin{array}{l}
\,\,\,\,\,k\biggl(\lambda_\alpha q_\alpha [q_\gamma]^{k-1}
c_1^{\beta[\gamma]_{k-1}|[\delta]_l}-\lambda_\beta q_\beta[q_\gamma]^{k-1}
c_1^{\alpha[\gamma]_{k-1}|[\delta]_l}\biggr)[p_\delta]^l+ \\
+l[q_\gamma]^k\biggl(c_1^{[\gamma]_k|\beta[\delta]_{l-1}}\lambda_\alpha
p_\alpha[p_\delta]^{l-1}-c_1^{[\gamma]_k|\alpha[\delta]_{l-1}}\lambda_\beta
p_\beta [p_\delta]^{l-1}\biggr)=0,\,\,\,3\le k,l<n,\,\,k+l\ge7.
\end{array}
\end{equation}
It follows from eq. (\ref{c23c}) that (see Appendix 1):
\begin{equation}\label{c23c1}
\begin{array}{l}
c_1^{[\alpha]_k|[\beta]_l}=0,\quad k\neq l,\quad k+l\neq n, \\
c_1^{[\alpha]_k|[\beta]_{n-k}}=c_1^{(k)}\varepsilon^{[\alpha]_k[\beta]_{n-k}},
\quad k,n-k\ge4,\quad 2k\neq n, \\
\varepsilon(c_1^{(k)})=n\,(mod\,\,2),\quad c_1^{(k)}=-c_1^{(n-k)}.
\end{array}
\end{equation}
Since $c_1^{(k)}$ are numbers they can be nonzero only for even $n$.

{\bf S t e p 4\quad} The coefficients in eq. (\ref{c2Jac}) in front of
$\partial_{\alpha_1}\partial_{\alpha_2}\partial_{\alpha_3}f_1(\xi)$ and
derivatives of $f_2(\xi)$ and $f_3(\xi)$ of orders $\ge3$.

The first three terms in eq. (\ref{c2Jac}) are equal to zero. The other
three give an equation of the following general structure:
\begin{equation}\label{c2Jac4}
c_1^{[\beta]_3|[\gamma]_{k+l-2}}[\partial]^kf_2[\partial]^lf_3+
c_1^{[\beta]_k|[\gamma]_{l+1}}[\partial]^kf_2[\partial]^lf_3+
c_1^{[\beta]_{k+1}|[\gamma]_l}[\partial]^kf_2[\partial]^lf_3=0.
\end {equation}

${\bf 4_a\quad}$ $k=l-1$, $4\le l<n$, $2l\neq n$,
$f_2(\xi)=\exp{(q_\beta\xi^\beta)}$.

In this case, the first two terms in eq. (\ref{c2Jac4}) are
equal to zero and we obtain:
\begin{equation}\label{c24a}
(\delta_{\alpha_1\beta}c_1^{\alpha_2\alpha_3[\beta]_{l-2}|[\gamma]_l}+
\delta_{\alpha_2\beta}c_1^{\alpha_3\alpha_1[\beta]_{l-2}|[\gamma]_l}+
\delta_{\alpha_3\beta}c_1^{\alpha_1\alpha_2[\beta]_{l-2}|[\gamma]_l})
\lambda_\beta q_\beta[q_\beta]^{l-2}=0.
\end {equation}

The general solution of this equation is (see Appendix 1)
$$
c_1^{[\beta]_l|[\gamma]_l} =0.
$$

Thus, the structure of all coefficients $c_1^{[\alpha]_k|[\beta]_l}$ is
determined for odd $n$.

${\bf 4_b\quad}$ $n=2m$ is even, $k=m-1$, $l=m$,
$f_2(\xi)=\exp{(q_\beta\xi^\beta)}$, $f_3(\xi)=\exp{(\xi^\gamma p_\gamma)}$.

Eq. (\ref{c2Jac4}) gives (with already established structure of the
coefficients $c_1^{[\alpha]_k|[\beta]_{2m-k}}$, $k\neq m$, see
formulas (\ref{c23c1})):
\begin{equation}\label{c24b}
\begin{array}{l}
\,\,\,\,\,3C^{m-2}_{2m-3}
c_1^{(3)}\varepsilon^{\alpha_1\alpha_2\alpha_3[\beta]_{m-2}[\gamma]_{m-1}}
\lambda_\delta q_\delta[q_\beta]^{m-2}p_\delta[p_\gamma]^{m-1}+ \\
+(-1)^mC_{m+1}^2c_1^{(m-1)}\biggl
(\varepsilon^{\alpha_1\alpha_2[\beta]_{m-1}[\gamma]_{m-1}}
\delta_{\alpha_3\gamma}+\hbox{cycle}(1,2,3)\biggr) [q_\beta]^{m-1}
\lambda_\gamma p_\gamma [p_\gamma]^{m-1}- \\
-C_m^2\biggl(c_1^{\alpha_1\alpha_2[\beta]_{m-2}|[\gamma]_m}
\delta_{\alpha_3\beta}+\hbox{cycle}(1,2,3)\biggr)
\lambda_\beta q_\beta[q_\beta]^{m-2}[p_\gamma]^m=0.
\end{array}
\end{equation}
It follows from eq. (\ref{c24b}) (see Appendix 1)
$$
c_1^{[\alpha]_m|[\beta]_m}=c_1^{(m)}\varepsilon^{[\alpha]_m[\beta]_m}.
$$
But then we have (compare with (\ref{c23c1}))
$$
c_1^{(m)}=-c_1^{(m)}=0,
$$
i.e.
$$
c_1^{[\alpha]_m|[\beta]_m}=0.
$$

${\bf 4_c\quad}$ The coefficients in eq. (\ref{c2Jac4}) in front of
$[\partial]^kf_2$, $[\partial]^lf_3$, $k+l=n-4$.

In this case the number of
indices for all coefficients $c_1^{[\alpha]_{k_1}|[\beta]_{k_2}}$ in eq.
(\ref{c2Jac}) is equal to $n$ ($k_1+k_2=n$). All these coefficients have the
structure
$$
c_1^{[\alpha]_k|[\beta]_{n-k}}=c_1^{(k)}\varepsilon{[\alpha]_k[\beta]_{n-k}},
\quad c_1^{(n-k)}=-c_1^{(k)},\quad c_1^{(m)}=0.
$$
Eq. (\ref{c2Jac4}) acquires the form
\begin{equation}\label{c24c}
\begin{array}{l}
\,\,\,\,\,3C^{k-1}_{n-3}c_1^{(3)}
\varepsilon^{\alpha_1\alpha_2\alpha_3[\beta]_{k-1}[\gamma]_{l-1}}
\lambda_\delta q_\delta[q_\beta]^{k-1}p_\delta[p_\gamma]^{l-1}- \\
-(-1)^kC_{n-k}^2c_1^{(k)}\biggl
(\varepsilon^{\alpha_1\alpha_2[\beta]_k[\gamma]_{l-1}}
\delta_{\alpha_3\gamma}+\hbox{cycle}(1,2,3)\biggr)
[q_\beta]^k\lambda_\gamma p_\gamma[p_\gamma]^{l-1}- \\
-C_{n-l}^2c_1^{(l)}\biggl(
\varepsilon^{\alpha_1\alpha_2[\beta]_{k-1}|[\gamma]_1}
\delta_{\alpha_3\beta}+\hbox{cycle}(1,2,3)\biggr)
\lambda_\beta q_\beta[q_\beta]^{k-1}[p_\gamma]^l=0.
\end{array}
\end{equation}
It follows from eq. (\ref{c24c}) (see Appendix 1)
$$
c_1^{(k)}=0, \forall k.
$$

Hence, we have established, that the general solution of eq. (\ref{c2Jac}) has
the following form  for $n\ge3$:
\begin{equation}\label{c2Gen1}
\begin {array}{l}
\,\,\,\,\,[f_1,f_2]_{*_1}(\xi)=\sigma_1f_1(\xi)
(\overleftarrow\partial_\alpha\lambda_\alpha\partial_\alpha)^3f_2(\xi)+
c_1^{(n)}f_1(\xi)[\overleftarrow\partial_\alpha]^n\varepsilon^{[\alpha]_n}
\varepsilon^{[\beta]_n}[\partial_\beta]^nf_2(\xi)+ \\
+c_{1n}\biggl(f_1(\xi)(1-{1\over2}\overleftarrow\partial_\alpha\xi^\alpha)
\varepsilon^{[\beta]_n}[\partial_\beta]^nf_2(\xi)-
f_1(\xi)[\overleftarrow\partial_\alpha]^n\varepsilon^{[\alpha]_n}
(1-{1\over2}\xi^\beta\partial_\beta)f_2(\xi)\biggr)+ \\
+d_St_1 (\xi|f_1,f_2),
\end{array}
\end{equation}
Remember that one may put $c_1^{(3)}=0$ according to (\ref{c2n3}) and
we shall accept this in what follows. By direct substitution of expression
(\ref{c2Gen1}) into eq. (\ref{c2Jac}) we see that it is satisfied
without any further restrictions on the coefficients.

Now let us check that the first three terms in the r.h.s. of eq.
(\ref{c2Gen1}) cannot be represented as $c_{triv} =d_St $.

To do this we should solve the equation
\begin{equation}\label{c3Triv}
\begin{array}{l}
\,\,\,\,\,\sigma f_1(\xi)
(\overleftarrow\partial\!\!_\alpha\lambda_\alpha\partial_\alpha)^3f_2(\xi)+
c^{(n)}f_1(\xi)[\overleftarrow\partial\!\!_\alpha]^n\varepsilon^{[\alpha]_n}
\varepsilon^{[\beta]_n}[\partial_\beta]^nf_2(\xi)+ \\
+c_n\biggl(f_1(\xi)(1-{1\over2}\overleftarrow\partial\!\!_\alpha\xi^\alpha)
\varepsilon^{[\beta]_n}[\partial_\beta]^nf_2(\xi)-
f_1(\xi)[\overleftarrow\partial\!\!_\alpha]^n\varepsilon^{[\alpha]_n}
(1-{1\over2}\xi^\beta\partial_\beta)f_2(\xi)\biggr)= \\
=d_St (\xi|f_1, f_2).
\end{array}
\end{equation}

However, we prefer to solve a more general equation
\begin{equation}\label{c3Triv'}
\hbox{l.h.s. of eq. (\ref{c3Triv})} =
T_\eta^{-1}(d_St(\xi|T_\eta f_1,T_\eta f_2)),
\end {equation}
where $T_\eta$ is an operator of changing the generators
(see eq. (\ref{Chan})).
Eq. (\ref{c3Triv'}) can be rewritten as:
\begin{equation}\label{c3Triv1}
\begin{array}{l}
\,\,\,\,\,\sigma f_1(\xi)
(\overleftarrow\partial\!\!_\alpha\lambda_\alpha\partial_\alpha)^3f_2(\xi)+
c^{(n)}f_1(\xi)[\overleftarrow\partial\!\!_\alpha]^n\varepsilon^{[\alpha]_n}
\varepsilon^{[\beta]_n}[\partial_\beta]^nf_2(\xi)+ \\
+c_n\biggl(f_1(\xi)(1-{1\over2}\overleftarrow\partial\!\!_\alpha\xi^\alpha)
\varepsilon^{[\beta]_n}[\partial_\beta]^nf_2(\xi)-
f_1(\xi)[\overleftarrow\partial\!\!_\alpha]^n\varepsilon^{[\alpha]_n}
(1-{1\over2}\xi^\beta\partial_\beta)f_2(\xi)\biggr)= \\
=\!\sum\limits_{k=0}^n\!\!t^{[\alpha]_k}(\zeta(\xi))
[\partial_{\zeta^\alpha}]^k\biggl(f_1(\xi)
\overleftarrow\partial\!\!_{\zeta^\gamma}\lambda_\gamma
\partial_{\zeta^\gamma}f_2(\xi)\biggr)\!\!-
\!\!f_1(\xi)\overleftarrow\partial\!\!_{\zeta^\gamma}\lambda_\gamma
\partial_{\zeta^\gamma}\!\biggl(\sum\limits_{k=0}^n\!t^{[\alpha]_k}
(\zeta(\xi))[\partial_{\zeta^\alpha}]^k f_2(\xi)\biggr)\!- \\
-\biggl(\sum\limits_{k=0}^nt^{[\alpha]_k}(\zeta(\xi))
[\partial_{\zeta^\alpha}]^kf_1(\xi)\biggr)
\overleftarrow\partial\!\!_{\zeta^\gamma}\lambda_\gamma\partial_{\zeta^\gamma}
f_2 (\xi)
\end{array}
\end{equation}
Remember that we consider $c_1^{(3)}=0$.

First of all, note that the highest order of the derivatives acting on both
functions in the l.h.s. of eq. (\ref{c3Triv1}) is $2n$,
while the highest order of the derivatives in the r.h.s. is $n+2$.
The equality $2n=n+2 $ can not be valid for $n\ge3$ so it
follows that $c^{(n)}=0$, $n\ge2$. Further, the term in the
l.h.s. of eq. (\ref{c3Triv1}) containing $f_1(\xi)$ without
derivatives contains also $[\partial_\alpha]^nf_2(\xi)$. At the same time
the term in the r.h.s. of this equation containing $f_1(\xi)$
without derivatives contains only the first derivative of the function
$f_2(\xi)$ from what it follows that $c_n=0$, $n\ge 2$.
As a result, eq. (\ref{c3Triv1}) for
$n\ge3$ accepts the form
\begin{equation}\label{c3Triv11}
\begin{array}{l}
\,\,\,\,\,\sigma f_1(\xi)
(\overleftarrow\partial\!\!_\alpha\lambda_\alpha\partial_\alpha)^3f_2(\xi)=
\sum\limits_{k=0}^nt^{[\alpha]_k}(\zeta(\xi))[\partial_{\zeta^\alpha}]^k
\biggl(f_1(\xi)\overleftarrow\partial\!\!_{\zeta^\gamma}\lambda_\gamma
\partial_{\zeta^\gamma}f_2(\xi)\biggr)- \\
-f_1(\xi)\overleftarrow\partial\!\!_{\zeta^\gamma}\lambda_\gamma
\partial_{\zeta^\gamma}\!\biggl
(\sum\limits_{k=0}^nt^{[\alpha]_k}(\zeta(\xi))[\partial_{\zeta^\alpha}]^k
f_2(\xi)\biggr)\!-\!\biggl(\sum\limits_{k=0}^nt^{[\alpha]_k}(\zeta(\xi))
[\partial_{\zeta^\alpha}]^kf_1(\xi)\biggr)\!
\overleftarrow\partial\!\!_{\zeta^\gamma}\lambda_\gamma\partial_{\zeta^\gamma}
f_2 (\xi)
\end{array}
\end{equation}
or, in the equivalent form,
\begin{equation}\label{c3Triv12}
\begin{array}{l}
\,\,\,\,\,\sigma f_1(\xi)
(\overleftarrow\partial\!\!_{\eta^\alpha}\lambda_\alpha
\partial_{\eta^\alpha})^3f_2 (\xi)
=\sum\limits_{k=0}^nt^{[\alpha]_k}(\xi)[\partial_\alpha]^k
\biggl(f_1(\xi)\overleftarrow\partial\!\!_\gamma\lambda_\gamma\partial_\gamma
f_2 (\xi) \biggr)- \\
-f_1(\xi)\overleftarrow\partial\!\!_\gamma\lambda_\gamma
\partial_\gamma\biggl(\sum\limits_{k=0}^nt^{[\alpha]_k}(\xi)
[\partial_\alpha]^kf_2(\xi)\biggr)-\biggl(\sum\limits_{k=0}^n
t^{[\alpha]_k}(\xi)[\partial_\alpha]^kf_1(\xi)\biggr)
\overleftarrow\partial\!\!_\gamma\lambda_\gamma\partial_\gamma f_2 (\xi).
\end{array}
\end{equation}

{\bf S t e p} ${\bf0'}$\quad The coefficient in eq. (\ref{c3Triv12}) in
front of $f_1(\xi)$ without derivatives gives:
$$
\partial_\gamma t^0(\xi)\lambda_\gamma\partial_\gamma f_2(\xi)=0,
\quad\quad\Longrightarrow\quad\quad t^0(\xi)=t_0=const.
$$

{\bf S t e p} ${\bf1'}$\quad The coefficients in eq. (\ref{c3Triv12}) at
$\partial_\alpha f_1\xi)$, $[\partial_\beta]^kf_2(\xi)$, $k\ge4$ give:
\begin{equation}\label{c31}
\partial_\alpha t^{[\beta]_k}(\xi)=0\quad\quad\Longrightarrow\quad\quad
t^{[\beta]_k}(\xi)=t^{[\beta]_k}=const,\quad k\ge4.
\end{equation}

{\bf S t e p} ${\bf2'}$\quad The coefficient in eq. (\ref{c3Triv12}) in
front of
$\partial_\alpha\partial_\beta f_1(\xi)$,
$f_2(\xi)=\exp{(\xi^\beta p_\beta)}$ provide:
\begin{equation}\label{c32}
\biggl(t^{\alpha[\gamma]_{k-1}}\delta_{\beta\gamma}-t^{\beta[\gamma]_k}
\delta_{\alpha\gamma}\biggr)\lambda_\gamma p_\gamma [p_\gamma] ^k=0,\quad
k\ge4.
\end {equation}
The general solution of eq. (\ref{c32}) is (see Appendix 1)
\begin{equation}\label{c32'}
t^{[\alpha]_k}=0,\quad 4\le k\le n-1,\quad
t^{[\alpha]_n}=t_n\varepsilon^{[\alpha]_n}.
\end {equation}
Eq. (\ref{c3Triv12}) now accepts the form
\begin{equation}\label{c3Triv12'}
\begin{array}{l}
\,\,\,\,\,\sigma f_1(\xi)
(\overleftarrow\partial\!\!_{\eta^\alpha}\lambda_\alpha
\partial_{\eta^\alpha})^3f_2(\xi)
=\sum\limits_{k=0}^3t^{[\alpha]_k}(\xi)[\partial_\alpha]^k
\biggl(f_1(\xi)\overleftarrow\partial\!\!_\gamma\lambda_\gamma\partial_\gamma
f_2 (\xi) \biggr)- \\
-f_1(\xi)\overleftarrow\partial\!\!_\gamma\lambda_\gamma
\partial_\gamma\biggl(\sum\limits_{k=0}^3t^{[\alpha]_k}(\xi)
[\partial_\alpha]^kf_2(\xi)\biggr)-\biggl(\sum\limits_{k=0}^3
t^{[\alpha]_k}(\xi)[\partial_\alpha]^kf_1(\xi)\biggr)
\overleftarrow\partial\!\!_\gamma\lambda_\gamma\partial_\gamma f_2 (\xi).
\end{array}
\end{equation}

{\bf S t e p} ${\bf3'}$\quad The coefficient in eq. (\ref{c3Triv12'}) in
front of
$[\partial_\alpha]^3f_1(\xi)$, $[\partial_\beta]^3f_2(\xi)$ provides:
$$
\sigma A^{[\alpha]_3}_{\gamma_1\gamma_2\gamma_3}\lambda_{\gamma_1}
\lambda_{\gamma_2}\lambda_{\gamma_3}A^{[\beta]_3}_{\gamma_1\gamma_2\gamma_3}
=0,\quad\quad\Longrightarrow\quad\quad\sigma=0.
$$

So, representations (\ref{c3Triv}) and (\ref{c3Triv'}) are exist only in the
case $\sigma\!=\!c^{(n)}\!=\!c_n\!=\!0$, i.e., in
particular, the second group of the Chevalley cogomology is
described by the parameters $\sigma$, $c^{(n)}$, $c_n$. For
completeness, we present the general solution of the equation
$$
d_St (\xi|f_1, f_2) =0.
$$
It has the form
$$
t(\xi|f)=t_0(1-{1\over2}\xi^\alpha\partial_\alpha)f(\xi)+
t(\xi)\overleftarrow\partial_\alpha\lambda_\alpha\partial_\alpha f(\xi)+
t_n\varepsilon^{[\alpha]_n}[\partial_\alpha]^nf(\xi).
$$

We have shown thus, that the general solution of eq. (\ref{c2Jac}) for
$n\ge3$ has form (\ref{c2Gen1}). Assuming, that the appropriate
similarity transformation eliminating the terms of the type
$d_St_1(\xi|f_1, f_2)$ from the expression for the $*$--commutator is
performed, we present the $*$--commutator for $n\ge3$ in the form:
\begin {equation}\label{c3h4}
\begin{array}{l}
\,\,\,\,\,C(\xi|f_1,f_2)=[f_1,f_2]_{M(\hbar\kappa_1\lambda)}(\xi)+
\hbar^2c_1^{(n)}f_1(\xi)[\overleftarrow\partial\!\!_\alpha]^n
\varepsilon^{[\alpha]_n} \varepsilon^{[\beta]_n}[\partial_\beta]^nf_2(\xi)+\\
+\hbar^2c_{1n}\biggl
(f_1(\xi)(1-{1\over2}\overleftarrow\partial\!\!_\alpha\xi^\alpha)
\varepsilon^{[\beta]_n}[\partial_\beta]^nf_2(\xi)-
f_1(\xi)[\overleftarrow\partial\!\!_\alpha]^n\varepsilon^{[\alpha]_n}
(1-{1\over2}\xi^\beta\partial_\beta)f_2(\xi)\biggr)+ \\
+\hbar^4c_2(\xi|f_1,f_2)+O(\hbar^6),
\end{array}
\end{equation}
$$c_2(\xi|f_1,f_2)=f_1(\xi)\sum_{k,l=0}^n[\overleftarrow\partial\!\!_\alpha]^k
c_2^{[\alpha]_k|[\beta]_l}(\xi)[\partial_\beta]^lf_2(\xi),
$$
$$
\varepsilon(c_2^{[\alpha]_k|[\beta]_l})=k+l\,(mod\,\,2),\quad
c_2^{[\beta]_l|[\alpha]_k}=-(-1)^{kl}c_2^{[\alpha]_k|[\beta]_l},
$$
where the Moyal bracket $[f_1,f_2]_{M(\hbar\kappa_1\lambda)}(\xi)$,
$\kappa_1=(6\sigma_1)^{1/2}$, is defined as follows:
\begin{equation}\label{c3M}
[f_1,f_2]_{M(\hbar\kappa_1\lambda)}(\xi)={1\over\hbar\kappa_1}f_1(\xi)
\sinh{(\hbar\kappa_1\overleftarrow\partial_\alpha\lambda_\alpha
\partial_\alpha)}f_2(\xi).
\end{equation}

The Moyal bracket (\ref{c3M}) satisfies the Jacobi identity (\ref{c1Jacob}),
(\ref{c1Jacob'}). Indeed, consider the associative $*$--product
$$
f_1*_{G(\hbar\kappa\lambda)}f_2(\xi)=f_1(\xi)
\exp{(\hbar\kappa\overleftarrow\partial\!\!_\alpha
\lambda_\alpha\partial_\alpha)} f_2 (\xi)
$$
for any complex $\kappa$. It is easy to verify, that it has the property
$$
f_2*_{G(\hbar\kappa\lambda)}f_1(\xi)=(-1)^{\varepsilon(f_1)\varepsilon(f_2)}
f_1(\xi)\exp{(-\hbar\kappa\overleftarrow\partial\!\!_\alpha
\lambda_\alpha\partial_\alpha)} f_2 (\xi).
$$
Therefore, the following representation is valid
$$
2f_1(\xi)\sinh{(\hbar\kappa\overleftarrow\partial\!\!_\alpha
\lambda_\alpha\partial_\alpha}) f_2 (\xi) =
f_1*_{G(\hbar\kappa\lambda)}f_2(\xi)-(-1)^{\varepsilon(f_1)\varepsilon(f_2)}
f_2*_{G(\hbar\kappa\lambda)}f_1(\xi),
$$
from which it follows that the Moyal bracket
$[f_1,f_2]_{M(\hbar\kappa\lambda)}(\xi)$ satisfies the Jacobi identity.

\section{Higher order deformations}

Now it is useful to point out, that in order to construct general solution
of eq. (\ref{c2Jac}), one needs equations only for the coefficients in front
of
$$
[\partial_\alpha]^kf_1(\xi),\quad k=0,1,2,
$$
and in front of
$$
[\partial_\alpha]^3f_1(\xi) [\partial_\alpha]^kf_2(\xi)
[\partial_\alpha]^lf_3,\quad k,l<n\quad(\hbox{for}\,\,n\ge4).
$$

Substitute representation (\ref{c3h4}) into the Jacobi identity
(\ref{c1Jacob}) which will be satisfied in the zeroth and first orders in
$\hbar^2$. In the $\hbar^4$ order we have:
\begin{equation}\label{c3Jac4}
\begin{array}{l}
\,\,\,\,\,\,(-1)^{\varepsilon(f_1)\varepsilon(f_3)}d_Sc_2(\xi|f_1,f_2,f_3)=\\
=2c_{1n}\sigma_1\biggl((-1)^{\varepsilon(f_1)\varepsilon(f_3)}
f_1(\xi)[\overleftarrow\partial\!\!_\alpha]^n\varepsilon^{[\alpha]_n}f_2(\xi)
(\overleftarrow\partial\!\!_\beta\lambda_\beta\partial_\beta)^3f_3(\xi)+
\hbox{cycle}(1,2,3)\biggr).
\end{array}
\end{equation}
In eq. (\ref{c3Jac4}) it was taken into account that $c_{1n}c_1^{(n)}=0$,
$\forall n$, since the factors $c_1^{(n)}$ are equal to zero for even $n$,
and the factors $c_{1n}$ are equal to zero for odd $n$.

Let $n\ge3$.

Consider the equations arising from vanishing of the coefficients
in front of $[\partial_\alpha]^kf_1(\xi)$, $k=0,1,2$, and in front of
$[\partial_\alpha]^3f_1(\xi)[\partial_\alpha]^kf_2(\xi)
[\partial_\alpha]^lf_3(\xi)$, $k,l<n$, in eq. (\ref{c3Jac4}). The r.h.s.
of eq. (\ref{c3Jac4}) does not give the contribution to these
equations, i.e., they coincide with the corresponding equations arising
from the solution of homogeneous equation (\ref{c2Jac}). As was already
noticed,  it is enough for complete definition of all coefficients of the
functional $c_2(\xi|f_1,f_2)$. Furthermore, the following equality will
be obviously satisfied
$$
d_Sc_2 (\xi|f_1, f_2, f_3) =0.
$$
Then it follows from eq. (\ref{c3Jac4}) that
$$
c_{1n}\sigma_1=0\quad
\Longrightarrow\quad\hbox{either}\,\,\,c_{1n}=0,\quad
\hbox{or}\,\,\,\sigma_1=0.
$$

By induction, we obtain that up to a similarity transformation, the
satisfying the conditions formulated in Sect. 2 nonsingular deformations of
the Poisson bracket on the Grassman algebra for $n\ge3$ are given by the
following expressions:

i) $n=2k$, $k\ge2$
\begin{equation}\label{c32k1}
[f_1,f_2]_{*^{(1)}}(\xi)=[f_1,f_2]_{M(\hbar\kappa\lambda)}(\xi),
\end{equation}
\begin{equation}\label{c32k2}
\begin{array}{l}
\,\,\,\,\,[f_1,f_2]_{*^{(2)}}(\xi)=[f_1,f_2]_{*_0}(\xi)+ \\
+\hbar^2c_n\biggl(f_1(\xi)(1-{1\over2}\overleftarrow
\partial\!\!_\alpha\xi^\alpha)\varepsilon^{[\beta]_n}[\partial_\beta]^n
f_2(\xi)-f_1(\xi)[\overleftarrow\partial\!\!_\alpha]^n\varepsilon^{[\alpha]_n}
(1-{1\over2}\xi^\beta\partial_\beta)f_2(\xi)\biggr).
\end{array}
\end{equation}

ii) $n=2k+1$, $k\ge1$
\begin{equation}\label{c32k+1}
[f_1,f_2]_{*^{(1)}}(\xi)=[f_1,f_2]_{M(\hbar\kappa\lambda)}(\xi)+\hbar^2
c^{(n)}f_1(\xi)[\overleftarrow\partial\!\!_\alpha]^n\varepsilon^{[\alpha]_n}
\varepsilon^{[\beta]_n}[\partial_\beta]^nf_2(\xi),
\end{equation}
\begin{equation}\label{c32k+12}
[f_1,f_2]_{*^{(2)}}(\xi)=[f_1,f_2]_{*_0}(\xi)+\hbar^2
c^{(n)}f_1(\xi)[\overleftarrow\partial\!\!_\alpha]^n\varepsilon^{[\alpha]_n}
\varepsilon^{[\beta]_n}[\partial_\beta]^nf_2(\xi).
\end {equation}
Certainly, $*^{(2)}$--commutator (\ref{c32k+12}) is a limiting
case (for $\kappa\rightarrow0$) of $*$--commutator (\ref{c32k+1}). The
parameters $\kappa$, $c_n$, $c^{(n)}$ can depend on $\hbar$. If one
requires that the functions $f(\xi)=const$ $*$--commute with any
function, then $c_n=0$.

In expressions (\ref{c32k1}) -- (\ref{c32k+1})
one can put $\kappa=\lambda_\alpha=1$, $c_n=0,\,1$ in the case
of the Grassman algebra ${\cal G}_{\bf C}$ and $\kappa=1,\,i$,
$\lambda_\alpha=\pm1$, $c_n=0,\pm1$ in the case of the Grassman algebra
${\cal G}_{\bf R}$. Besides, in eq. (\ref{c32k+12})
we may set $c^{(n)}=0,\,1$ in the case of the Grassman algebra
${\cal G}_{\bf C}$ and $c^{(n)}=0,\,\pm1$ in the case of the Grassman algebra
${\cal G}_{\bf R}$.

Indeed, after the similarity transformation of $*$--commutators
(\ref{c32k1}) -- (\ref{c32k+1}) by the operator $T=T^{(\mu)}$,
\begin{equation}\label{Tm}
T^{(\mu)}f(\xi)={1\over\mu^2}f(\mu\xi),\quad\mu\in K,
\end{equation}
the transformed $*$--commutators will preserve structure (\ref{c32k1}) --
(\ref{c32k+1}) (in particular, the zeroth approximation in $ \hbar^2 $ does
not vary), but with new parameters:
\begin{equation}\label{CP}
\kappa\quad\rightarrow\quad\kappa\mu^2,\quad\quad
c^{(n)}\quad\rightarrow\quad c^{(n)}\mu^{2(n-1)},\quad\quad
c_n\quad\rightarrow\quad c_n\mu^{n-2}.
\end{equation}
It is possible to fix $\mu$ imposing $\kappa\mu^2=1$ in eqs. (\ref{c32k1})
and (\ref{c32k+1}) and $c_n\mu^{n-2}=1$ for $c_n\neq0$ in eq.
(\ref{c32k2}) for the Grassman algebra ${\cal G}_{\bf C}$.
In the case of the Grassman algebra ${\cal G}_{\bf R}$ it is possible to
satisfy $\kappa\mu^2=1$ in formulas (\ref{c32k1}) and
(\ref{c32k+1}) for $\sigma_1>0$. If $\sigma_1<0$, then $\kappa$ is pure
imaginary and it is possible to fix $\mu$ from the condition
$\kappa\mu^2=i$. Note that in this case the Moyal bracket cannot be
presented as a commutator in the algebra with associative product.
The parameter $c_n$ in formula (\ref{c32k2}) can be normalized to
$+1,\,0,\,-1$ for $c_n>0,\,=0,\,<0$, respectively. If the parameter
$\kappa$ is fixed in eq. (\ref{c32k+1}), then we already do not
have a possibility to change the parameter $c^{(n)}$ and it remains
arbitrary.  However, in formula (\ref{c32k+12}) parameter $c^{(n)}$
can be normalized to $0,\,1$ for $c^{(n)}=0,\,\neq0$ in the case of
the Grassman algebra ${\cal G}_{\bf C}$ and to $+1,\,0,\,-1$ for
$c^{(n)}>0,\,=0,\,<0$ in the case of the Grassman algebra
${\cal G}_{\bf R}$.

Consider now the remaining cases $n=1$ and $n=2$.

i) $n=1$
$$
f=a+b\xi,
$$
$$
[f_1,f_2]_{*_0}(\xi)=\lambda f_1(\xi)\overleftarrow\partial\!\!\partial
f_1(\xi)=\lambda b_1b_2,\quad\lambda =\pm1,
$$
$$
\begin{array}{l}
\,\,\,\,\,[f_1,f_2]_*(\xi)=\lambda
f_1(\xi)\overleftarrow\partial\!\!\partial f_1(\xi)+ \\
+\hbar^2c_1(\xi)(-1)^{\varepsilon(f_1)}\biggl(f_1(\xi)\partial f_2(\xi)+
f_1(\xi)\overleftarrow\partial\!\!f_2(\xi)\biggr)+
\hbar^2c_2(\xi)f_1(\xi)\overleftarrow\partial\!\!\partial f_1(\xi),
\end{array}
$$
$$
\varepsilon(c_1(\xi))=1\Rightarrow c_1(\xi)=c_1\xi,\quad
\varepsilon (c_2(\xi))=0\Rightarrow c_2(\xi)=c_2,\quad c_1,c_2=const.
$$
After the similarity transformation generated by
$\xi\rightarrow\xi'=(1+\hbar^2c_2/\lambda)^{-1/2}\xi$, we obtain the
following expression for the $*$--commutator:
$$
\begin{array}{l}
\,\,\,\,\,\,[f_1,f_2]_*(\xi)=\lambda
f_1(\xi)\overleftarrow\partial\!\!\partial f_1(\xi)+
\hbar^2c_1\xi(-1)^{\varepsilon(f_1)}\biggl(f_1(\xi)\partial
f_2(\xi)+f_1(\xi)\overleftarrow\partial\!\!f_2(\xi)\biggr)= \\
=\lambda b_1b_2+\hbar^2c_1\xi(a_1b_2-b_1a_2).
\end{array}
$$

Take $f_1(\xi)=a_1$, $\varepsilon(f_1)=0$, $f_{2,3}(\xi)=b_{2,3}\xi$,
$\varepsilon(f_{2,3})=1$ and consider the Jacobi identity (\ref{c1Jacob}) for
these functions:
$$
0=[f_1,[f_2,f_3]_*]_*+[f_2,[f_3,f_1]_*]_*-[f_3,[f_1,f_2]_*]_*=
-2\hbar^2\lambda c_1a_1b_2b_3, \quad \Longrightarrow\quad c_1=0.
$$

Thus, the nontrivial deformation of the nonsingular Poisson bracket are absent
in the case of the Grassman algebra with one generator.

ii) $n=2$

Using the previous results (see, in particular, eqs. (\ref{c3Jac4}) and
(\ref{c2Gen12})), we obtain:

Up to the similarity transformations, the possible deformations of the
nonsingular Poisson bracket on the Grassman algebra with two generators are
$$
\begin{array}{l}
\,\,\,\,\,[f_1,f_2]_*(\xi)=[f_1,f_2]_{*_0}(\xi)+ \\
+\hbar^2c_2\biggl(
f_1(\xi)(1-{1\over2}\overleftarrow\partial\!\!_\alpha\xi^\alpha)
\varepsilon^{[\beta]_2}[\partial_\beta]^2f_2(\xi)-
f_1(\xi)[\overleftarrow\partial\!\!_\alpha]^2\varepsilon^{[\alpha]_2}
(1-{1\over2}\xi^\beta\partial_\beta)f_2(\xi)\biggr),
\end{array}
$$
$c_2$ can depend on $\hbar$, i.e., are given in fact by expression
(\ref{c32k2}). Note, that formally formulas (\ref{c32k1}) --
(\ref{c32k+12}) are valid for all $n$, since the Moyal bracket for $n=1,2$
reduces to the Poisson bracket, and the second terms in the r.h.s. of eqs.
(\ref{c32k+1}) and (\ref{c32k+12}) are proportional to the
Poisson bracket for $n=1$.

Thus, we have shown that an arbitrary $*$--commutator (satisfying conditions
formulated in Sect. 2) can be reduced to form (\ref{c32k1}) --
(\ref{c32k+12}) with the help of the similarity transformations. Is
it also possible to reduce expressions (\ref{c32k1}) -- (\ref{c32k+12})
to the Poisson bracket by the similarity transformation? The answer
to this question is in negative. It is enough to consider the case of
the Grassman algebra ${\cal G}_{\bf C}$. Present the operator $T$ generating
the similarity transformation as
$$
T=T'(1+\hbar^2t_1+O(\hbar^4)),\quad T'=T\big|_{\hbar=0}
$$
($T'$ is nonsingular operator). Due to boundary condition (\ref{BC}) the
similarity transformation generated by the operator $T'$ should retain
the zeroth order in $\hbar^2$ approximation to the $*$-commutator to be the
Poisson bracket (maybe with a different symplectic metric):
$$
T'{}^{-1}\biggl([T'f_1,T'f_2]_{*_0}\biggr)(\xi)=[f_1,f_2]_{*_{0'}}(\xi)=
T_{\eta'}^{-1}\biggl([T_{\eta'}f_1,T_{\eta'}f_2]_{*_0}\biggr)(\xi),
$$
where $T_{\eta'}$ is an operator of the change of generator, reducing
the Poisson bracket $*_{0'}$ into the canonical form $*_0$. The operator
$T''=T_{\eta'}^{-1}T'$ satisfies the equation
\begin{equation}\label{IP}
T''[f_1,f_2]_{*_0}(\xi)=[T''f_1,T''f_2]_{*_0}(\xi).
\end{equation}
The general form of such operators is found in Appendix 2:
$$
T''=T^{(\mu)}T_\eta^{(c)}T^{(n)},
$$
$$
T_\eta^{(c)}f(\xi)=f(\eta(\xi)),\quad\quad
T^{(n)}=(1+t_n\varepsilon^{[\alpha]_n}[\partial_\alpha]^n),\quad
t_n=const,\quad\varepsilon(t_n)=n(mod\,2),
$$
$\varepsilon(\eta^\alpha)=1$, the change $\xi^\alpha$ $\rightarrow$
$\eta^\alpha(\xi)$ is canonical, the operator $T^{(\mu)}$ is determined by
(\ref{Tm}). Then the operator $T'$ is given by
$$
T'=T^{(\mu)}T_\zeta T^{(n)},
$$
where $T_\zeta$ is an operator of some change of the generators.

It is easy to verify that the similarity transformation generated by
the operator $T^{(n)}$ leaves invariant not only the $*_0$--commutator
(the Poisson bracket), but also the exact $*$--commutator given by formulas
(\ref{c32k1}) -- (\ref{c32k+1}). The similarity transformation
generated by the operator $T^{(\mu)}$ results only in multiplication of
parameters in eqs. (\ref{c32k1}) -- (\ref{c32k+1}) by nonzero
factors (see eq. (\ref{CP})). Thus, the possibility of reducing the
$*$--commutator to the $*_0$--commutator in the first $\hbar^2$ order
depends on the validity of eq. (\ref{c3Triv'}), which, as has been shown, has
no solutions for the nonzero parameters $\kappa$, $c^{(n)}$, $c_n$.

We have finally shown, that the following theorem is valid

{\bf T h e o r e m}

Up to similarity transformation, the satisfying the conditions formulated in
Sect. 2 deformations of the nonsingular Poisson bracket on the Grassman
algebra are given by expressions (\ref{c32k1}) --
(\ref{c32k+12}) and $\kappa=1$, $c_n=0,1$ in the case of the Grassman algebra
${\cal G}_{\bf C}$ and $\kappa=1,\,i$, $c_n=0,\,\pm1$ in the case of the
Grassman algebra ${\cal G}_{\bf R}$; $c^{(n)}$ is an arbitrary complex
number in the case of the algebra ${\cal G}_{\bf C}$ and an arbitrary real
number in the case of the algebra ${\cal G}_{\bf R}$ in formula (\ref{c32k+1}),
and $c^{(n)}=0,\,1$ in the case of the algebra ${\cal G}_{\bf C}$ and
$c^{(n)}=0,\,\pm1$ in the case of the algebra ${\cal G}_{\bf R}$ in formula
(\ref{c32k+12}). Furthermore note, that $*$--commutators (\ref{c32k1})
-- (\ref{c32k+12}) with different values of the parameter $\hbar$ can also
be connected by the similarity transformation generated by the operator
$T^{(\mu)}$ with suitable $\mu$.

{\bf Acknowledgments}

The author thanks I.Tipunin for useful discussions and RFBR,
contract 99--01--00980 and school--contract 00--15--96566, for support.

\setcounter{equation}{0}
\def\theequation{A.\arabic{equation}}

\section*{Appendix 1}

In this appendix the solutions of some equations from the main text are
presented.

{\bf Equations (\ref{0a}), (\ref{1a2}) and (\ref{1a1}).}

Consider equation
\begin{equation}\label{Cog}
d\Omega(x,\xi)=0,\quad d=\xi^\alpha\lambda_\alpha
{\partial\over\partial x^\alpha},\quad d^2=0,
\end {equation}
where $x^\alpha$ are the set of even variables (generators),
$x^\alpha x^\beta-x^\beta x^\alpha=x^\alpha\xi^\beta-\xi^\beta x^\alpha=0 $.
Introduce the operators
$$
\gamma=x^\alpha\lambda_\alpha{\partial\over\partial\xi^\alpha},\quad
N=x^\alpha{\partial\over\partial x^\alpha}+
\xi^\alpha{\partial\over\partial\xi^\alpha},
$$
generating the algebra
$$
\gamma^2=0,\quad d\gamma+\gamma d=0,\quad [d,N]=[\gamma,N]=0.
$$
It is easy to see, that if the function $\Omega(x,\xi)$ satisfies the eq.
(\ref{Cog}) and the operator $N^{-1}$ is defined on $\Omega(x,\xi)$  (it
is necessary to have $\Omega(0,0)=0$ for this), then the function
$\Omega(x,\xi)$ can be presented as
$$
\Omega(x,\xi)=d\Xi(x,\xi),\quad \Xi(x,\xi)=\gamma{1\over N}\Omega(x,\xi),
$$
and the general solution of eq. (\ref{Cog}) is
\begin{equation}\label{Omeg}
\Omega(x,\xi)=d\Xi(x,\xi)+const.
\end{equation}
Let $\Omega(x,\xi)=x^\alpha c^\alpha(\xi)$ also be solution of eq. (\ref{Cog}).
Then it follows from eqs. (\ref{Cog}) and (\ref{Omeg}) that
$$
\lambda_\alpha\partial_\alpha c^\beta(\xi)+\lambda_\beta\partial_\beta
c^\alpha(\xi)=0,\quad c^\alpha(\xi)=\lambda_\alpha\partial_\alpha c (\xi),
$$
It gives the solutions of eqs. (\ref{0a}) and (\ref{1a2}). Similarly,
choosing $\Omega(x,\xi)=x^\alpha x^\beta c^{\alpha\beta}(\xi)$, we obtain
the solution of eq. (\ref{1a1}).

{\bf Equations (\ref{c20b}), (\ref{c23a1}) and (\ref{c32}).}

All these equations can be reduced to the form
\begin{equation}\label{c42}
\biggl(a^{\alpha[\gamma]_{k-1}}(\xi)\delta_{\beta\gamma}-
a^{\beta[\gamma]_k}(\xi)\delta_{\alpha\gamma}\biggr)\lambda_\gamma
p_\gamma[p_\gamma]^{k-1}=0,\quad k\ge2,
\end {equation}
where $a^{[\alpha]_k}$ $=$ $c_1^{0|[\alpha]_k}$, $c_1^{[\alpha]_k}$ or
$t^{[\alpha]_k}$. Acting by the operator
$\lambda_\beta\partial/\partial p_\beta$ on eq. (\ref{c42}) one gets:
$$
(n-k)a^{[\alpha]_k}=0\quad\Longrightarrow\quad a^{[\alpha]_k}=0,\quad
2\le k\le n-1.
$$
Note that $a^{[\alpha]_n}=a_n\varepsilon^{[\alpha]_n}$ satisfies
eq. (\ref {c42}) identically.

{\bf Equation (\ref{c22a})}
\begin{equation}\label{c42a}
\begin {array}{l}
\,\,\,\,\,\biggl(\delta_{\gamma\beta}\lambda_\beta
c_1^{\alpha\sigma|[\delta]_l}-\delta_{\gamma\alpha}\lambda_\alpha
c_1^{\beta\sigma|[\delta]_l}-\delta_{\sigma\beta}\lambda_\beta
c_1^{\alpha\gamma|[\delta]_l}+\delta_{\sigma\alpha}\lambda_\alpha
c_1^{\beta\gamma|[\delta]_l}\biggr)[p_\delta]^l= \\
=l\biggl(\delta_{\alpha\delta}c_1^{\gamma\sigma|\beta[\delta]_{l-1}}-
\delta_{\beta\delta}c_1^{\gamma\sigma|\alpha[\delta]_{l-1}}-
\delta_{\gamma\delta}c_1^{\alpha\beta|\sigma[\delta]_{l-1}}+
\delta_{\sigma\delta}c_1^{\alpha\beta|\gamma[\delta]_{l-1}}\biggr)
\lambda_\delta p_\delta[p_\delta]^{l-1}.
\end{array}
\end{equation}

Acting by the operator $\lambda_\alpha\partial/\partial_\alpha$
on eq. (\ref{c42a}) we obtain:
\begin{equation}\label{c42a1}
\begin{array}{l}
\,\,\,\,\,(n-l)c_1^{\gamma\sigma|\beta[\delta]_{l-1}}[p_\delta]^{l-1}=
\lambda_\beta\biggl(\delta_{\beta\sigma}\Delta^{\gamma|[\delta]_{l-1}}-
\delta_{\beta\gamma}\Delta^{\sigma|[\delta]_{l-1}}\biggr)[p_\delta]^{l-1}+\\
+(l-1)\biggl(\delta_{\sigma\delta}\Delta^{\beta|\gamma[\delta]_{l-2}}-
\delta_{\gamma\delta}\Delta^{\beta|\sigma[\delta]_{l-2}}\biggr)
\lambda_\delta p_\delta[p_\delta]^{l-2},\quad\Delta^{\alpha|[\beta]_{l-1}}
\equiv\lambda_\gamma\delta_{\gamma\sigma}
c_1^{\alpha\gamma|\sigma[\beta]_{l-1}}.
\end{array}
\end{equation}
Now, acting by the operator $p_\sigma\lambda_\beta\partial/\partial_\beta$
on eq. (\ref{c42a1}) one gets:
$$
(l-2)p_\gamma\Delta^{[p_\delta]_{l-2}}[p_\delta]^{l-2}=0,\quad
\Delta^{[p_\delta]_{l-2}}\equiv\lambda_\alpha\delta_{\alpha\beta}
\Delta^{\alpha|\beta[\delta]_{l-2}},
$$
from where it follows that $\Delta^{[\delta]_{l-2}}$ can be nonzero
only for $l=2 $ or $l-2=n$. Since $l\le n$, we get that
$\Delta^{[\delta]_{l-2}}=0$ for $l\neq2$.

Multiplying eq. (\ref{c42a1}) by $\lambda_\beta\delta_{\beta\sigma}$
one obtains:
\begin{equation}\label{c42a2}
\Delta^{\gamma|[p_\delta]_{l-1}}[p_\delta]^{l-1}+
\Delta^{\delta|\gamma[p_\delta]_{l-2}}p_\delta[p_\delta]^{l-2}=
\Delta^{[p_\delta]_{l-2}}\lambda_\gamma p_\gamma[p_\delta]^{l-2}.
\end {equation}
Applying  the operator $\lambda_\gamma\partial/\partial_\gamma$ to eq.
(\ref{c42a2}) we find that $\Delta^{[\delta]_{l-2}}=0$ also if $l=2$. Eq.
(\ref{c42a2}) acquires the form:
\begin{equation}\label{c42a2'}
\Delta^{\gamma|[p_\delta]_{l-1}}[p_\delta]^{l-1}+
\Delta^{\delta|\gamma[p_\delta]_{l-2}}p_\delta[p_\delta]^{l-2}=0.
\end {equation}
Differentiating eq. (\ref{c42a2'}) over $p_\sigma$ we obtain
$$
l\Delta^{\gamma|\sigma[\delta]_{l-2}}[p_\delta]^{l-2}=
(l-2)(\Delta^{\delta|\gamma\sigma[\delta]_{l-3}}p_\delta[p_\delta]^{l-3}+
(\Delta^{\gamma|\sigma[\delta]_{l-2}}-
\Delta^{\sigma|\gamma[\delta]_{l-2}})[p_\delta]^{l-2},
$$
from where it follows that $\Delta^{\alpha|[\alpha]_{l-1}}$ are totally
antisymmetric in all indices:
$$
\Delta^{\alpha|[\alpha]_{l-1}}=\Delta^{[\alpha]_l}.
$$
Multiplying eq. (\ref{c42a1}) for $l<n$ by $p_\beta $ we finally find
\begin{equation}\label{c42a3}
c_1^{\alpha\beta|[\delta]_l}[p_\delta]^l=(\delta_{\alpha\delta}
a_1^{\beta[\delta]_{l-1}}-\delta_{\beta\delta}
a_1^{\alpha[\delta]_{l-1}})\lambda_\delta p_\delta
[p_\delta]^{l-1},\,\,\,a_1^{\alpha[\alpha]_{l-1}}\equiv-
{l\over n-l}\Delta^{[\alpha]_l},\,\,\,2\le l<n.
\end {equation}
The r.h.s. of eq. (\ref{c42a}) vanishes for $l=n$ (to
see this, it is necessary to use the equality
$c_1^{\alpha\beta|[\gamma]_n}=c_1^{\alpha\beta}\varepsilon^{[\gamma]_n}$ and
the identity $\varepsilon^{\beta[\delta]_{n-1}}p_\alpha[p_\delta]^{n-1}=
(1/n)\delta_{\alpha\beta}\varepsilon^{[\delta]_n}[p_\delta]^n$). Multiplying
eq. (\ref{c42a}) by $\lambda_\alpha\delta_{\alpha\gamma}$:
$$
(2-n)c_1^{\beta\sigma|[\delta]_n}[p_\delta]^n=0,
$$
we obtain $c_1^{\beta\sigma|[\delta]_n}=0$. Note that representation
(\ref{c42a3}) is valid also for $l=n$.

{\bf Equation (\ref{c23b33})}
\begin{equation}\label{c43b33}
\begin{array}{l}
\,\,\,\,\,\delta_{\alpha\alpha_1}\lambda_\alpha
c_1^{\beta\alpha_2\alpha_3|\beta_1\beta_2\beta_3}+
\delta_{\alpha\alpha_2}\lambda_\alpha
c_1^{\beta\alpha_3\alpha_1|\beta_1\beta_2\beta_3}+
\delta_{\alpha\alpha_3}\lambda_\alpha
c_1^{\beta\alpha_1\alpha_2|\beta_1\beta_2\beta_3}- \\
-\delta_{\beta\alpha_1}\lambda_\beta
c_1^{\alpha\alpha_2\alpha_3|\beta_1\beta_2\beta_3}-
\delta_{\beta\alpha_2}\lambda_\beta
c_1^{\alpha\alpha_3\alpha_1|\beta_1\beta_2\beta_3}-
\delta_{\beta\alpha_3}\lambda_\beta
c_1^{\alpha\alpha_1\alpha_2|\beta_1\beta_2\beta_3}+ \\
+\delta_{\alpha\beta_1}\lambda_\alpha
c_1^{\alpha_1\alpha_2\alpha_3|\beta\beta_2\beta_3}+\delta_{\alpha\beta_2}
\lambda_\alpha c_1^{\alpha_1\alpha_2\alpha_3|\beta\beta_3\beta_1}+
\delta_{\alpha\beta_3} \lambda_\alpha
c_1^{\alpha_1\alpha_2\alpha_3|\beta\beta_1\beta_2}- \\
-\delta_{\beta\beta_1}\lambda_\beta
c_1^{\alpha_1\alpha_2\alpha_3|\alpha\beta_2\beta_3}-
\delta_{\beta\beta_2}\lambda_\beta
c_1^{\alpha_1\alpha_2\alpha_3|\alpha\beta_3\beta_1}-
\delta_{\beta\beta_3}\lambda_\beta
c_1^{\alpha_1\alpha_2\alpha_3|\alpha\beta_1\beta_2}=0.
\end{array}
\end{equation}

Multiply eq. (\ref{c43b33}) by
$\lambda_\alpha\delta_{\alpha\alpha_1}$ (and
then replace $\beta$ by $\alpha_1$):
\begin{equation}\label{c43b331}
\begin{array}{l}
\,\,\,\,\,(n-2)c_1^{\alpha_1\alpha_2\alpha_3|\beta_1\beta_2\beta_3}-
[c_1^{\alpha_2\alpha_3\alpha_1|\beta_1\beta_2\beta_3}-
c_1^{\alpha_2\alpha_3\beta_1|\beta_2\beta_3\alpha_1}+
c_1^{\alpha_2\alpha_3\beta_2|\beta_3\alpha_1\beta_1}-
c_1^{\alpha_2\alpha_3\beta_3|\alpha_1\beta_1\beta_2}]= \\
=\lambda_{\alpha_1}\biggl(\delta_{\alpha_1\beta_1}
\Sigma^{\alpha_2\alpha_3|\beta_2\beta_3}+
\delta_{\alpha_1\beta_2}\Sigma^{\alpha_2\alpha_3|\beta_3\beta_1}+
\delta_{\alpha_1\beta_3}\Sigma^{\alpha_2\alpha_3|\beta_1\beta_2}\biggr),
\end{array}
\end{equation}
$$
\Sigma^{\alpha_1\alpha_2|\beta_1\beta_2}\equiv\lambda_\gamma
c_1^{\alpha_1\alpha_2\gamma|\gamma\beta_1\beta_2}=
\Sigma^{\beta_1\beta_2|\alpha_1\alpha_2}.
$$
The symmetry properties of the tensor
$\Sigma^{\alpha_1\alpha_2|\beta_1\beta_2}$ follow from antisymmetry
condition (\ref{c2Asym}).

Multiplying eq. (\ref{c43b331}) by
$\lambda_{\alpha_3}\delta_{\alpha_3\beta_3}$ one has:
\begin{equation}\label{c43b332}
(n-1)\Sigma^{\alpha_1\alpha_2|\beta_1\beta_2}-
[\Sigma^{\alpha_1\alpha_2|\beta_1\beta_2}+
\Sigma^{\alpha_1\beta_1|\beta_2\alpha_2}+
\Sigma^{\alpha_1\beta_2|\alpha_2\beta_1}]=
\lambda_{\alpha_1}\biggl(\delta_{\alpha_1\beta_2}\Sigma^{\alpha_2|\beta_1}-
\delta_{\alpha_1\beta_1}\Sigma^{\alpha_2|\beta_2}\biggr),
\end {equation}
$$
\Sigma^{\alpha|\beta}\equiv\lambda_\gamma\Sigma^{\alpha\gamma|\gamma\beta}=
\Sigma^{\beta |\alpha}.
$$
The multiplication of eq. (\ref{c43b331}) by
$\lambda_{\alpha_1}\delta_{\alpha_1\alpha_2}$ gives
\begin{equation}\label{c43b333}
\Sigma^{\alpha\beta_1|\beta_2\beta_3}+
\Sigma^{\alpha\beta_2|\beta_3\beta_1}+
\Sigma^{\alpha\beta_3|\beta_1\beta_2}=0
\end {equation}
Finally, we multiply eq. (\ref{c43b332}) by
$\lambda_{\alpha_2}\delta_{\alpha_2\beta_1}$:
\begin{equation}\label{c43b334}
\Sigma^{\alpha|\beta}={1\over n}\lambda_\alpha\delta_{\alpha\beta}\Sigma,
\end {equation}
$$
\Sigma=\lambda_\gamma\Sigma^{\gamma|\gamma},\quad
\varepsilon(\Sigma)=0.
$$
Substituting expression (\ref{c43b334}) into eq. (\ref{c43b332}) and
using condition (\ref{c43b333}), we obtain
\begin{equation}\label{c43b335}
\Sigma^{\alpha_1\alpha_2|\beta_1\beta_2}=
-{1\over n(n-1)}\lambda_{\alpha_1}\lambda_{\alpha_2}
(\delta_{\alpha_1\beta_1}\delta_{\alpha_2\beta_2}-\delta_{\alpha_1\beta_2}
\delta_{\alpha_2\beta_1})\Sigma.
\end{equation}
Represent $c_1^{\alpha_1\alpha_2\alpha_3|\beta_1\beta_2\beta_3}$ as
$$
\begin{array}{l}
\,\,\,\,\,c_1^{\alpha_1\alpha_2\alpha_3|\beta_1\beta_2\beta_3}=-
{\lambda_{\alpha_1}\lambda_{\alpha_2}\lambda_{\alpha_3}
\Sigma\over n(n-1)(n-2)}\biggl(\delta_{\alpha_1\beta_1}
\delta_{\alpha_2\beta_2}\delta_{\alpha_3\beta_3}+
\delta_{\alpha_1\beta_3}\delta_{\alpha_2\beta_1}\delta_{\alpha_3\beta_2}+
\delta_{\alpha_1\beta_2}\delta_{\alpha_2\beta_3}\delta_{\alpha_3\beta_1}- \\
-\delta_{\alpha_1\beta_2}\delta_{\alpha_2\beta_1}\delta_{\alpha_3\beta_3}-
\delta_{\alpha_1\beta_3}\delta_{\alpha_2\beta_2}\delta_{\alpha_3\beta_1}-
\delta_{\alpha_1\beta_1}\delta_{\alpha_2\beta_3}\delta_{\alpha_3\beta_2}
\biggr)+\tilde{c}_1^{\alpha_1\alpha_2\alpha_3|\beta_1\beta_2\beta_3}.
\end {array}
$$
Substituting this representation into eq. (\ref{c43b331}) we have
\begin{equation}\label{c43b336}
(n-2) \tilde c_1^{\alpha_1\alpha_2\alpha_3|\beta_1\beta_2\beta_3}=
\tilde c_1^{\alpha_2\alpha_3\alpha_1|\beta_1\beta_2\beta_3}-
\tilde c_1^{\alpha_2\alpha_3\beta_1|\beta_2\beta_3\alpha_1}+
\tilde c_1^{\alpha_2\alpha_3\beta_2|\beta_3\alpha_1\beta_1}-
\tilde c_1^{\alpha_2\alpha_3\beta_3|\alpha_1\beta_1\beta_2}.
\end {equation}
The r.h.s. of eq. (\ref{c43b336}) is antisymmetric under the
permutation of the indices $\alpha_1$ and $\beta_1$,
therefore, its l.h.s., the tensor
$\tilde c_1^{\alpha_1\alpha_2\alpha_3|\beta_1\beta_2\beta_3}$, changes a
sign under the permutation of $\alpha_1$ and $\beta_1$. Hence
the tensor $\tilde c_1^{\alpha_1\alpha_2\alpha_3|\beta_1\beta_2\beta_3}$ is
totally antisymmetric in all indices. In particular,
$$
\tilde c_1^{\alpha_1\alpha_2\alpha_3|\beta_1\beta_2\beta_3}=-
\tilde c_1^{\beta_1\beta_2\beta_3|\alpha_1\alpha_2\alpha_3}.
$$
However, from the condition (\ref{c2Asym}) it follows
$\tilde c_1^{\alpha_1\alpha_2\alpha_3|\beta_1\beta_2\beta_3}=
\tilde c_1^{\beta_1\beta_2\beta_3|\alpha_1\alpha_2\alpha_3}$, i.e.
$ \tilde c_1^{\alpha_1\alpha_2\alpha_3|\beta_1\beta_2\beta_3}=0$.
Thus, we finally obtain
$$
c_1^{\alpha_1\alpha_2\alpha_3|\beta_1\beta_2\beta_3}=-
{\lambda_{\alpha_1}\lambda_{\alpha_2}\lambda_{\alpha_3}
\Sigma\over n(n-1)(n-2)}\biggl
(\delta_{\alpha_1\beta_1}\delta_{\alpha_2\beta_2}\delta_{\alpha_3\beta_3}+
\delta_{\alpha_1\beta_3}\delta_{\alpha_2\beta_1}\delta_{\alpha_3\beta_2}+
\delta_{\alpha_1\beta_2}\delta_{\alpha_2\beta_3}\delta_{\alpha_3\beta_1}-
$$
$$
-\delta_{\alpha_1\beta_2}\delta_{\alpha_2\beta_1}\delta_{\alpha_3\beta_3}-
\delta_{\alpha_1\beta_3}\delta_{\alpha_2\beta_2}\delta_{\alpha_3\beta_1}-
\delta_{\alpha_1\beta_1}\delta_{\alpha_2\beta_3}\delta_{\alpha_3\beta_2}
\biggr)
$$

{\bf Equation (\ref{c23c})}
\begin{equation}\label{c43c}
\begin{array}{l}
\,\,\,\,\,\,k\biggl(\lambda_\alpha q_\alpha[q_\gamma]^{k-1}
c_1^{\beta[\gamma]_{k-1}|[\delta]_l}-\lambda_\beta q_\beta[q_\gamma]^{k-1}
c_1^{\alpha[\gamma]_{k-1}|[\delta]_l}\biggr)[p_\delta]^l+ \\
+l[q_\gamma]^k\biggl(c_1^{[\gamma]_k|\beta[\delta]_{l-1}}\lambda_\alpha
p_\alpha[p_\delta]^{l-1}-c_1^{[\gamma]_k|\alpha[\delta]_{l-1}}\lambda_\beta
p_\beta[p_\delta]^{l-1}\biggr)=0,\quad 3\le k,l<n.
\end{array}
\end{equation}
Acting on eq. (\ref{c43c}) by the operator
$\lambda_\beta\partial/\partial q_\beta$ we get:
\begin{equation}\label{c43c1}
(n-l)[q_\gamma]^kc_1^{[\gamma]_k|\alpha[\delta]_{l-1}}+
kq_\delta[q_\gamma]^{k-1}c_1^{\alpha[\gamma]_{k-1}|\delta[\delta]_{l-1}}+
(-1)^kk\lambda_\alpha q_\alpha[q_\gamma]^{k-1}
\Delta^{[\gamma]_{k-1}|[\delta]_{l-1}},
\end {equation}
$$
\Delta^{[\gamma]_{k-1}|[\delta]_{l-1}}\equiv\lambda_\gamma
c_1^{[\gamma]_{k-1}\gamma|\gamma[\delta]_{l-1}}.
$$
Act on eq. (\ref{c43c1}) by the operator
$\lambda_\alpha\partial/\partial q_\alpha$ one has:
$$
(k-l)\Delta^{[\gamma]_{k-1}|[\delta]_{l-1}}=0,\quad\Longrightarrow\quad
\Delta^{[\gamma]_{k-1}|[\delta]_{l-1}}=0,\quad k\neq l.
$$
Take $k\neq l$ in eq. (\ref{c43c1}). Acting on this equation by the operator
$\partial/\partial q_\alpha$ we obtain:
$$
[q_\gamma]^{k-1}[(n-l)c_1^{\alpha[\gamma]_{k-1}|\beta[\delta]_{l-1}}+
c_1^{\beta[\gamma]_{k-1}|\alpha[\delta]_{l-1}}]=(k-1)q_\delta[q_\delta]^{k-2}
c_1^{\beta\alpha[\gamma]_{k-2}|\delta[\delta]_{l-1}}.
$$
Combining this expression with the expression with permuted indices
$\alpha$ and $\beta$ one gets:
$$
c_1^{\alpha[\gamma]_{k-1}|\beta[\delta]_{l-1}}+
c_1^{\beta[\gamma]_{k-1}|\alpha[\delta]_{l-1}}=0,
$$
from where it follows that $c_1^{[\alpha]_k|[\beta]_l}$ is totally
antisymmetric in all indices for $k\neq l$. Using this fact in eq.
(\ref{c43c1}), we obtain
$$
(n-k-l)c_1^{[\alpha]_k|[\beta]_l}=0.
$$
Thus, we finally have
$$
c_1^{[\alpha]_k|[\beta]_l}=0,\quad 3\le k,l<n,\quad k\neq l,\quad
k+l\neq n,
$$
$$
c_1^{[\alpha]_k|[\beta]_{n-k}}=c_1^{(k)}
\varepsilon^{[\alpha]_k|[\beta]_{n-k}},\quad k,n-k\ge4\quad 2k\neq n.
$$

{\bf Equation (\ref{c24a})}
\begin{equation}\label{c44a}
(\delta_{\alpha_1\beta}c_1^{\alpha_2\alpha_3[\beta]_{l-2}|[\gamma]_l}+
\delta_{\alpha_2\beta}c_1^{\alpha_3\alpha_1[\beta]_{l-2}|[\gamma]_l}+
\delta_{\alpha_3\beta}c_1^{\alpha_1\alpha_2[\beta]_{l-2}|[\gamma]_l})
\lambda_\beta q_\beta [q_\beta] ^ {l-2} =0.
\end{equation}

Acting on eq. (\ref{c44a}) by the operator $\lambda_{\alpha_3}
\partial/\partial q_{\alpha_3}$ we obtain (after cancelation of $(n-2)$)
$$
c_1^{\alpha_1\alpha_2[\beta]_{l-2}|[\gamma]_l}=0.
$$

{\bf Equation (\ref{c24b})}
\begin{equation}\label{c44b}
\begin{array}{l}
\,\,\,\,\,3C^{m-2}_{2m-3}c_1^{(3)}
\varepsilon^{\alpha_1\alpha_2\alpha_3[\beta]_{m-2}[\gamma]_{m-1}}
\lambda_\delta q_\delta[q_\beta]^{m-2}p_\delta[p_\gamma]^{m-1}+ \\
+(-1)^mC_{m+1}^2c_1^{(m-1)}\biggl
(\varepsilon^{\alpha_1\alpha_2[\beta]_{m-1}[\gamma]_{m-1}}
\delta_{\alpha_3\gamma}+\hbox{cycle}(1,2,3)\biggr)[q_\beta]^{m-1}
\lambda_\gamma p_\gamma[p_\gamma]^{m-1}- \\
-C_m^2\biggl(c_1^{\alpha_1\alpha_2[\beta]_{m-2}|[\gamma]_m}
\delta_{\alpha_3\beta}+\hbox{cycle}(1,2,3)\biggr)
\lambda_\beta q_\beta[q_\beta]^{m-2}[p_\gamma]^m=0.
\end{array}
\end{equation}

Acting on eq. (\ref{c44b}) by the operator
$\lambda_{\alpha_3}\partial/\partial q_{\alpha_3}$ one has:
$$
\biggl((-1)^m3C^{m-2}_{2m-3}c_1^{(3)}+(m-1)C_{m+1}^2c_1^{(m-1)}\biggr)
\varepsilon^{\alpha_1\alpha_2[\beta]_{m-2}[\gamma]_m}[q_\beta]^{m-2}
[p_\gamma]^m-
$$
$$
-mC_m^2c_1^{\alpha_1\alpha_2[\beta]_{m-2}|[\gamma]_m}[q_\beta]^{m-2}
[p_\gamma]^m=0,
$$
from where it follows
$$
c_1^{[\alpha]_m|[\beta]_m}=c_1^{(m)}\varepsilon^{[\alpha]_m[\beta]_m},\quad
c_1^{(m)}={1\over mC_m^2}\biggl((-1)^m3C^{m-2}_{2m-3}c_1^{(3)}+
C_{m+1}^2c_1^{(m-1)}\biggr).
$$

{\bf Equation (\ref{c24c}), $n=2m$}.
\begin{equation}\label{c44c}
\begin{array}{l}
\,\,\,\,\,3C^{k-1}_{n-3}c_1^{(3)}
\varepsilon^{\alpha_1\alpha_2\alpha_3[\beta]_{k-1}[\gamma]_{l-1}}
\lambda_\delta q_\delta[q_\beta]^{k-1}p_\delta[p_\gamma]^{l-1}- \\
-(-1)^kC_{n-k}^2c_1^{(k)}\biggl
(\varepsilon^{\alpha_1\alpha_2[\beta]_k[\gamma]_{l-1}}
\delta_{\alpha_3\gamma}+\hbox{cycle}(1,2,3)\biggr)
[q_\beta] ^k\lambda_\gamma p_\gamma[p_\gamma]^{l-1}- \\
-C_{n-l}^2c_1^{(l)}\biggl(
\varepsilon^{\alpha_1\alpha_2[\beta]_{k-1}|[\gamma]_1}
\delta_{\alpha_3\beta}+\hbox{cycle}(1,2,3)\biggr)
\lambda_\beta q_\beta [q_\beta]^{k-1}[p_\gamma]^l=0.
\end{array}
\end{equation}
Acting on eq. (\ref{c44c}) by the operator
$\lambda_{\alpha_3}\partial/\partial q_{\alpha_3}$ we have:
\begin{equation}\label{c44c1}
3C^{k-1}_{n-3}c_1^{(3)}-(-1)^kkC_{n-k}^2c_1^{(k)}+
(-1)^klC_{n-l}^2c_1^{(l)}=0.
\end{equation}
Eq. (\ref{c44c1}) is simplified, if one rewrites it in terms of $d_k $:
$$
d_k\equiv(-1)^kk!(n-k)!c_1^{(k)},\quad d_{n-k}=-d_k,
$$
\begin{equation}\label{c44c2}
d_{3}+d_{k}-d_{k+1}=0.
\end {equation}
It follows from eq. (\ref{c44c2})
$$
d_k=(k-2)d_3.
$$
Remembering that $0=c_1^{(m)}=d_m$, we obtain
$$
d_3=0\quad\Longrightarrow\quad c_1^{(k)}=0\quad\forall k.
$$

\section*{Appendix 2}

In this Appendix we find the general form of the operators $T''$ generating
similarity transformations, which leave invariant the Poisson bracket on
the Grassman algebra ${\cal G}_{\bf C}$. The operators $T''$ satisfy
the condition (see (\ref{IP}))
\begin{equation}\label{c4Inv}
T''[f_1,f_2]_{*_0}(\xi)=[T''f_1,T''f_2]_{*_0}(\xi).
\end{equation}

We use the momentum representation for the operator $T''$:
$$
T''=\sum_{k=0}^nt^{[\alpha]_k}(\xi)[\partial_\alpha]^k,\quad
\varepsilon(t^{[\alpha] _k})=k\,(mod\,\,2),
$$
In such terms eq. (\ref{c4Inv}) reads:
\begin{equation}\label{c4Inv'}
\begin{array}{l}
\,\,\,\,\,\,\sum\limits_{k=0}^nt^{[\alpha]_k}(\xi)[\partial_\alpha]^k\biggl
(\partial_\gamma f_1(\xi)\lambda_\gamma\partial_\gamma f_2(\xi)\biggr)= \\
=\biggl(\sum\limits_{k_1=0}^n\partial_\gamma t^{[\alpha]_{k_1}}(\xi)
[\partial_\alpha]^{k_1}+\sum\limits_{k_1=0}^nt^{[\alpha]_{k_1}}(\xi)
[\partial_\alpha]^{k_1}\partial_\gamma\biggr)f_1(\xi)\lambda_\gamma\times\\
\times\,\,\biggl(\sum\limits_{k_2=0}^n
\partial_\gamma t^{[\beta]_{k_2}}(\xi)[\partial_\beta]^{k_2}+
\sum\limits_{k_2=0}^nt^{[\beta]_{k_2}}(\xi)[\partial_\beta]^{k_2}
\partial_\gamma\biggr)f_2(\xi).
\end{array}
\end{equation}

{\bf S t e p ${\bf0}$\quad} The coefficient in eq. (\ref{c4Inv'}) in front
of $f_1(\xi)$ gives:
$$
\partial_\gamma
t^0(\xi)\lambda_\gamma\partial_\gamma\biggl(\sum\limits_{k=0}^n
t^{[\beta]_k}(\xi)[\partial_\beta]^kf_2(\xi)\biggr)=0\quad
\Longrightarrow\quad t^0 (\xi) =t_0=const.
$$
From nonsingularity of the operator $T''$ it follows that $t_0\neq0$.
Present the operator $T''$ as
$$
T''=T^{(\mu)}T^{(1)},
$$
\begin{equation}\label{c401}
T^{(\mu)}f(\xi)={1\over\mu^2}f(\mu\xi),
\end{equation}
where $\mu=t_0^{1/2}$. It is easy to verify, that the similarity
transformation generated by the operator $T^{(\mu)}$ (and by its inverse),
leaves the $*_0$--commutator invariant. Hence, the operator $T^{(1)}$ and its
coefficients (for which we retain previous notation $t^{[\alpha]_k}$) satisfy
eq. (\ref{c4Inv}) and eq. (\ref{c4Inv'}), respectively. Furthermore, the
operator $T^{(1)}$ has $t_0=1 $.

{\bf S t e p ${\bf 1}$\quad} The coefficient in (\ref{c4Inv'}) in front of
$\partial_\alpha f_1(\xi)$, $\partial_\beta f_2(\xi)$ ($t_0=1$) gives:
$$
\{\eta^\alpha(\xi),\eta^\beta(\xi)\}=\lambda_\alpha\delta_{\alpha\beta}=
\{\xi^\alpha,\xi^\beta\},
$$
where the notation $\eta^\alpha(\xi)\equiv t^\alpha(\xi)+\xi^\alpha$ is
used. Thus, the transformation of generators $\xi^\alpha$ $\rightarrow$
$\eta^\alpha(\xi)$ is canonical. Let us present the operator $T''$ as
$$
T''=T^{(\mu)}T_\eta^{(c)}T^{(2)},
$$
where the operator $T^{(\mu)} $ is defined by eq. (\ref{c401}) and the
operator $T_\eta^{(c)}$ describes a canonical transformation of the
generators: $T_\eta^{(c)}f(\xi)=f(\eta(\xi))$. It is easy to verify,
that the similarity transformation generated by the operator $T_\eta^{(c)}$,
leaves the $*_0$--commutator invariant. Hence, the similarity transformation
generated by the operator $T^{(2)}$ also leaves the $*_0$--commutator
invariant. Besides the operator $T^{(2)}$ has $t^0(\xi)=1$, $t^\alpha(\xi)=0$
(we again retain the previous notation $t^{[\alpha]_k}$ for the
coefficients of the operator $T^{(2)}$). In what follows we consider eq.
(\ref{c4Inv'}) as an equation for the coefficients of the operator
$T^{(2)}$.

{\bf S t e p ${\bf1'}$\quad} The coefficients in (\ref{c4Inv'}) in front of
$\partial_\alpha f_1(\xi)$, $[\partial_\beta]^kf_2(\xi)$, $k\ge2$, give:
$$
\partial_\alpha t^{[\beta]_k}(\xi)=0\quad\quad\Longrightarrow\quad\quad
t^{[\alpha]_k}(\xi)=t^{[\alpha]_k}=const.
$$

{\bf S t e p ${\bf2}$\quad} The coefficient in (\ref{c4Inv'}) in front of
$\partial_\alpha\partial_\beta f_1(\xi)$,
$f_2(\xi)=\exp{(\xi^\alpha p_\alpha)}$, gives:
$$
(\delta_{\alpha\gamma}t^{\beta[\gamma]_{k-1}}-\delta_{\beta\gamma}
t^{\alpha[\gamma]_{k-1}})\lambda_\gamma p_\gamma[p_\gamma]^{k-1}=0.
$$
The general solution of this equation (see the solution of eq. (\ref{c42}))
is
$$
t^{[\alpha]_k}=0,\quad2\le k\le n-1,\quad\quad
t^{[\alpha]_n}=t_n\varepsilon^{[\alpha]_n},
$$
$t_n\neq0 $ only at even $n$, from where it follows
$$
T^{(2)}\equiv T^{(n)}=1+t_n\varepsilon^{[\alpha]_n}[\partial_\alpha]^n.
$$
Thus, the general form of the operators $T''$ is:
$$
T''=T^{(\mu)}T_\eta^{(c)}T^{(n)}.
$$

\end{document}